\pdfoutput=1

\documentclass[conference]{IEEEtran}
\IEEEoverridecommandlockouts

\usepackage{tikz}

\usepackage{xcolor}
\usepackage{dsfont}
\usepackage{numprint}
\usepackage{subfig}
\usepackage{caption}
\usepackage[export]{adjustbox}

\usepackage{url}
\usepackage{setspace}
\usepackage{tabularx}  

\usepackage{indentfirst}
\usepackage{amsthm,amsmath}
\usepackage[ruled,lined,linesnumbered]{algorithm2e}  
\usepackage{algorithmic}
\usepackage{booktabs}
\usepackage{nicematrix}
\usepackage{mathrsfs}
\usepackage{graphicx}
\usepackage{pgfplots}
\pgfplotsset{compat=1.17}

\usepackage{color}
\usepackage{verbatim}
\usepackage{textcomp}

\usepackage{bbold}

\usepackage[utf8]{inputenc}
\usepackage{array}
\usepackage{makecell}
\usepackage{wrapfig}

\usepackage{multirow}

\usepackage{soul}
\usepackage{balance}

\usepackage[english]{babel}
\usepackage{blindtext}
\usepackage[normalem]{ulem}

\newcommand{\Paragraph}[1]{~\vspace*{-0.8\baselineskip}\\{\textbf{#1}}}

\newcommand{\Cli}{\mathcal{P}}
\newcommand{\B}{\mathcal{B}}
\newcommand{\LL}{\mathcal{L}_}
\newcommand{\xx}{\langle x\rangle}

\graphicspath{{./plot/}}

\def\BibTeX{{\rm B\kern-.05em{\sc i\kern-.025em b}\kern-.08em
    T\kern-.1667em\lower.7ex\hbox{E}\kern-.125emX}}

\begin{document}

\title{Private Data Valuation and Fair Payment in Data Marketplaces}
\author{
    \IEEEauthorblockN{Zhihua Tian\IEEEauthorrefmark{3}, Jian Liu\IEEEauthorrefmark{1}, Jingyu Li\IEEEauthorrefmark{1}, Xinle Cao\IEEEauthorrefmark{1}, Ruoxi Jia\IEEEauthorrefmark{2}, Jun Kong\IEEEauthorrefmark{4}, Mengdi Liu\IEEEauthorrefmark{4}, Kui Ren\IEEEauthorrefmark{1}}\\
    \IEEEauthorblockA{\IEEEauthorrefmark{3}Zhejiang University and ZJU-Hangzhou Global Scientific and Technological Innovation Center,\\
    \IEEEauthorrefmark{1}Zhejiang University, \IEEEauthorrefmark{2}Virginia Tech, \IEEEauthorrefmark{4} Zhejing Big Data Exchange Center\\
    \{zhihuat, liujian2411, jingyuli, xinle, kuiren\}@zju.edu.cn, ruoxijia@vt.edu, \{kongj, liumd\}@zjdex.com}
}






\maketitle
\begin{abstract}
Data valuation is an essential task in a data marketplace. It aims at fairly compensating data owners for their contribution. There is increasing recognition in the machine learning community that the Shapley value---a foundational profit-sharing scheme in cooperative game theory---has major potential to value data, because it uniquely satisfies basic properties for fair credit allocation and
has been shown to be able to identify data sources that are useful or harmful to model performance. However, calculating the Shapley value requires accessing original data sources. It still remains an open question how to design a real-world data marketplace that takes advantage of the Shapley value-based data pricing while protecting privacy and allowing fair payments.

In this paper, we propose the {\em first} prototype of a data marketplace that values data sources based on the Shapley value in a privacy-preserving manner and at the same time ensures fair payments. Our approach is enabled by a suite of innovations on both algorithm and system design. We firstly propose a Shapley value calculation algorithm that can be efficiently implemented via multiparty computation (MPC) circuits. The key idea is to learn a performance predictor that can directly predict model performance corresponding to an input dataset without performing actual training. We further optimize the MPC circuit design based on the structure of the performance predictor. We further incorporate fair payment into the MPC circuit to guarantee that the data that the buyer pays for is exactly the same as the one that has been valuated. Our experimental results show that the proposed new data valuation algorithm is as effective as the original expensive one. Furthermore, the customized MPC protocol is efficient and scalable.
\end{abstract}


\begin{IEEEkeywords}
Data marketplace, Fair payment, Data valuation
\end{IEEEkeywords}





\section{Introduction}

{\em Data, a new factor of production,} is now treated equally as land, labour, capital and technology.
As a result, data commoditization has become an emerging trend: 
data buyers seek to purchase high-quality data, and data owners attempt to maximize data monetization.
A {\em data marketplace} is a platform where users can buy and sell data. 
It facilitates data owners marketing, managing, and selling their data; 
it also enables data buyers to browse, compare, and purchase data.

Data marketplaces can be roughly classified into three categories based on what they sell and their {\em pricing mechanisms}: data-based, query-based, and model-based. 
Data marketplaces with data-based pricing sell data to buyers directly, for example Dawex~\cite{Dawex}, Datarade~\cite{Datarade}, Quandl~\cite{Quandl} and Bloomberg~\cite{Bloomberg}.
Such marketplaces give data owners limited control over their data usages, and data buyers have to purchase the whole dataset even if they are only interested in particular information.
Data marketplaces with query-based pricing charge buyers and compensate data owners on a per-query basis, e.g., Google Bigquery~\cite{BigQuery}.
This type of marketplace can usually only allow simple queries; in particular, they are unable to support widely used machine learning (ML) based data analytics.

To address these issues, model-based pricing~\cite{DBLP:journals/pvldb/JiaDWHGLZSS19} has recently been proposed as an alternative to the two types of pricing mechanisms mentioned above.
A model-based pricing mechanism valuates each data source according to its contribution to training a given ML model. For instance, a collection of multiple data sources is used jointly to train an ML model, which achieves certain performance, say classification accuracy 0.9. The model-based pricing mechanism assigns price among all data sources, so that each source receives a fair share for its contribution towards achieving the 0.9
accuracy. Most works have focused on leveraging the Shapley value as the metric to quantify the contribution of individual data sources
~\cite{DBLP:conf/aistats/JiaDWHHGLZSS19,DBLP:journals/pvldb/JiaDWHGLZSS19,DBLP:conf/icml/GhorbaniZ19}. Particularly, it \emph{uniquely} satisfies several natural properties of an equitable data pricing mechanism: 1) it requires that the the contribution of a data source is sensitive only to how the model performance responds to the presence of the source; 2) it ensures that the contribution of data to predicting multiple test inputs is equal to the sum of contributions to predicting each individual test input; and 3) it distributes the full yield of an ML pipeline. Aside from the theoretical soundness, it has been demonstrated in~\cite{DBLP:conf/cvpr/Jia0SXDK00S21,DBLP:conf/icml/GhorbaniZ19} that the Shapley value can reliably differentiate between high-quality and low-quality data.

Despite the appealing properties of the Shapley value, none of the existing data marketplace platforms supports its adaption.
Some platforms only allow buyers to request samples to \emph{qualitatively} understand the quality of the data that they intend to buy~\cite{Datarade, Quandl}. Indeed, deploying SV-based data pricing mechanisms into the real-world data marketplace requires systematic innovations to tackle the following dilemmas.

\newenvironment{packeditemize}{
\begin{list}{$\bullet$}{
\setlength{\labelwidth}{8pt}
\setlength{\itemsep}{0pt}
\setlength{\leftmargin}{\labelwidth}
\addtolength{\leftmargin}{\labelsep}
\setlength{\parindent}{0pt}
\setlength{\listparindent}{\parindent}
\setlength{\parsep}{0pt}
\setlength{\topsep}{3pt}}}{\end{list}}

\begin{packeditemize}
    \item {\em Who performs the data valuation?} Calculation of the Shapley value must access all individual data sources.  
    If the buyer performs the data valuation, the buyer has to access the data and as a result, the buyer no longer needs to buy the data.
    If the data owners perform the valuation, they have to access the valuation algorithm and therefore, they can potentially manipulate their data to maximize the profit. 
    \item {\em Who delivers first?}
    If the buyer delivers the payment first, a data owner might refuse to deliver the data or deliver inferior data.
    If the data owners deliver data first, the buyer might refuse to pay.
\end{packeditemize}
Firstly, introducing a broker between the buyer and the data owners does not resolve these two dilemmas. The broker could potentially collude with either the buyer or the data owners. Secondly, it may seem natural to resort to cryptographic techniques to resolve the first dilemma.
For example, to address the first dilemma, data owners and the buyer can jointly run data valuation via {\em multiparty computation} (MPC)~\cite{goldreich1998secure}, such that the buyer can get SV without seeing the data. However, even computing SV in plaintext could be expensive because it requires retraining models on different subsets of data; calculating them via MPC remains an open problem. 
As for the second dilemma, one natural idea is to leverage
{\em zero-knowledge contingent payment} (ZKCP)~\cite{10.1007/978-3-030-88428-4_31, DBLP:conf/ccs/CampanelliGGN17,DBLP:conf/ccs/NguyenAA20,DBLP:conf/ccs/LiYHMGZZSLW21}, which allows fair payment of digital goods and payments over a blockchain.
However, simply applying ZKCP cannot ensure that data owners use the same data for both MPC-based data valuation and ZKCP.
As a result, a malicious data owner could use high-quality data to get a high SV in MPC-based data valuation, and then deliver inferior data in ZKCP. \emph{Overall, while a combination of MPC and ZKCP provides a natural pathway to approach the dilemmas, it suffers significant computational issues and also exposes new vulnerabilities for a malicious user to game the system.}

\textbf{Our contributions.}
In this paper, we propose the {\em first} framework to solve the aforementioned dilemmas, thus enabling private data valuation and fair payment in a data marketplace. The proposed framework includes a suite of sophisticated designs.

First of all, we {\bf design an MPC-friendly algorithm for SV calculation}.
Given $n$ data sources, the exact SV calculation requires repeatedly training an ML model over every possible combination of data sources, leading to $\mathcal{O}(2^n)$ complexity.
Although existing work leverages Monte Carlo-based algorithms to approximate the SV by training over polynomially many randomly chosen subsets, training ML models is still too heavy for MPC. 
Inspired by the recent advance in active learning, we propose to learn an ML \textit{performance predictor} that directly predicts the performance of the model corresponding to a training set without performing actual training. 
Note that in existing data marketplaces~\cite{Datarade, Quandl}, sellers often pre-share a small portion of representative samples of their data to facilitate buyers' assessment of the quality and relevance of the data. Our performance predictor is learned in a supervised manner with pre-shared samples.
The learned performance predictor can be implemented in MPC to support calculation of the SV of each dataset in a way that does not reveal the original data.
Compared with the existing data marketplaces in which the buyers can only acquire a qualitative understanding of a dataset from pre-shared samples, our system makes a better use of pre-shared data by offering a quantitative evaluation of the dataset's contribution.

Second, we {\bf optimize the MPC circuit design for the performance predictor} to further improve efficiency. 
It is well known that the complexity of MPC increases \textit{quadratically} with the number of parties.
Therefore, directly running MPC for data valuation in data marketplaces cannot support a large number of data owners.
Recall that we directly predict the performance of each combination of data with a parameter model.
The prediction process consists of an encoding phase that encodes each data sample into an embedding and a mapping phase that aggregates the embeddings to a vector and maps the vector to the performance. To that end, we optimize the MPC circuit such that it only requires all data owners to get involved in the mapping phase. The encoding phase can be achieved by data owners running the two-party computation (2PC) separately with the buyer.
Since the mapping phase consists of only a few fully connected layers, which is efficient to implement via MPC, the cost of data valuation is nearly unaffected by the number of data owners.

Finally, instead of using ZKCP, we {\bf innovatively incorporate fair payment into our MPC circuit}. 
In more detail, each data owner inputs an encryption key $k$ together with her data $D$ to MPC.
Other than SV, the MPC protocol will also outputs $E(k, D)$ and $H(k)$ to the buyer, where $E()$ is a symmetric-key encryption scheme and $H()$ is a cryptographic hash function. 
If the buyer decides to buy the data, he issues a {\em hash-locked transaction}~\cite{Hashlock} on a blockchain, requiring the data owner to post the preimage of $H(k)$ to the blockchain to redeem the payment.
Most cryptocurrencies like Bitcoin and Ethereum allow a payer to make a payment by specifying a condition that needs to be met in order for the money to be redeemed by the payee.
A hash-locked transaction is such a condition, where a payment can be redeemed by presenting a preimage of a hash value. 
Then, the data owner can simply post $k$ to the blockchain to redeem the payment; meanwhile, the buyer can decrypt $E(k, D)$ and obtain the purchased data.
The MPC protocol ensures that the data being encrypted is exactly the data being valuated. 
We are the first to adopt the idea of incorporating fair payment into the MPC circuit to address the fair payment problem in data marketplaces.

We provide a full implementation of our data valuation algorithms as well as  the MPC circuits. 
Our experimental results show that the proposed data valuation algorithms are as effective as the original expensive one.
Furthermore, the customized MPC protocol is efficient and scalable.

\textbf{Paper Organization.}
The rest of the paper is organized as follows. 
Section~\ref{sec:overview} presents the overview of our method.
We provide our data valuation algorithms in Section~\ref{sec:data_evaluation}.
The customized MPC protocol is presented in Section~\ref{ssec:optimization_mpc_circuit} and we show how to incorporate fair payment into it in Section~\ref{ssec:fair_payment}.
Section~\ref{sec:experiments_for_utility} presents the experiments for evaluating the effectiveness of the data valuation algorithms as well as the efficiency of the MPC circuits. Section~\ref{sec:related_work} presents related work and Section~\ref{sec:conclusion} concludes the paper.


\textbf{Notations.} We summarize the frequently used notations in Table~\ref{tab:notations} for the convenience of readers.
\begin{table}[ht]
\small
\centering
\caption{Summary of notations}
\begin{tabular*}{8cm}{r|l}
\toprule
\textbf{Notation} & \textbf{Description} \\ 
\midrule
$\Cli$ & data owner \\ 
$\B$ & the buyer \\ 
$u$ & data utility function \\
$C$ & coalition of datasets \\
$SV$ & Shapley values of all data owners \\ 
$v_{sv}$ & Shapley value of one data owner \\ 
$v_{loo}$ & Leave-one-out value of one data owner \\ 
$G_f$ & feature extractor \\
$f_{DS}$ & DeepSets model \\
$\mathcal{A}$ & learning algorithm\\
$\LL{tr}$ & training set \\ 
$\LL{val}$ & validation set \\ 
$\LL{pub}$ & public set \\ 
$M$ & size of the train dataset\\
$\mathcal{S}_n$ & a $n$ secret sharing scheme \\
$\xx$ & shared variable \\ 
$\xx_n$ & shares calculated via n secret sharing \\ 
$\xx_{i/n}$ & individual share of $\xx_n$ held by player $i$ \\
\bottomrule
\end{tabular*}
\label{tab:notations}
\vspace{-3mm}
\end{table}

\begin{figure*}[t]
    \centering
    \includegraphics[width=\textwidth]{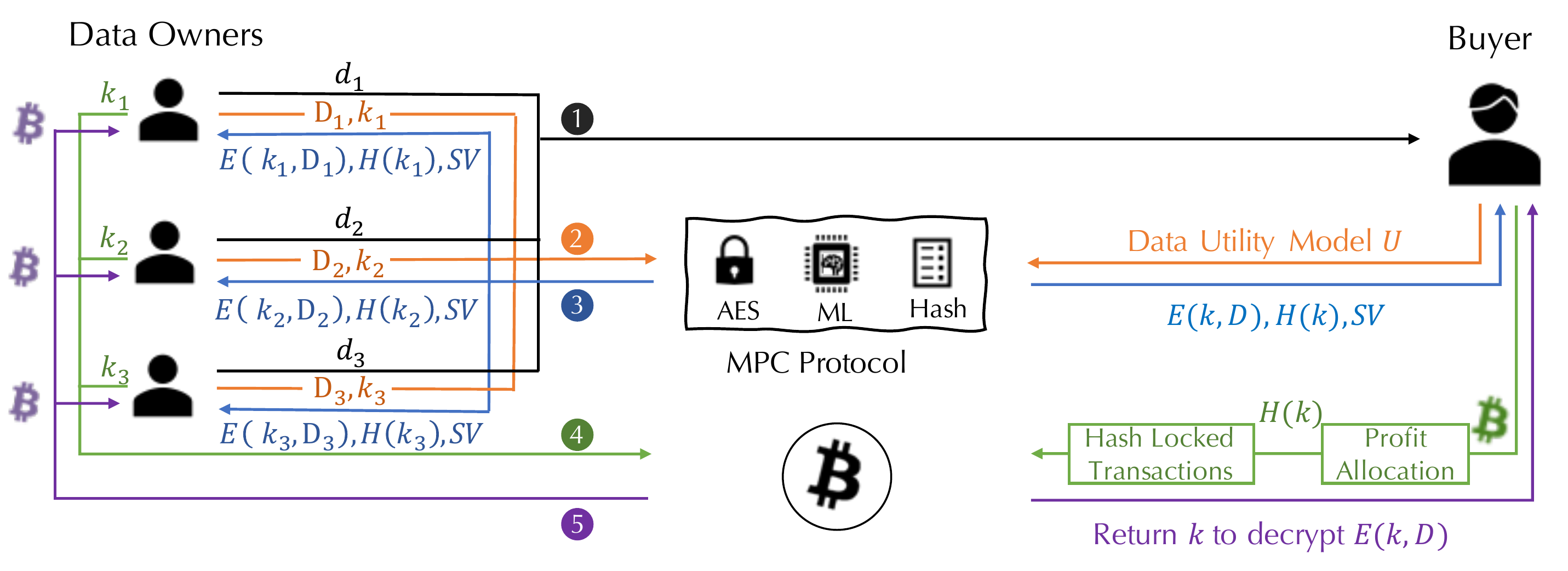}
    \caption{Overview of our proposed framework addressing both private data valuation and fair payment.}
    \label{fig:overview}
\end{figure*}

\section{Overview}
\label{sec:overview}
We consider the common setting of a data marketplace: 
$N$ data owners $\Cli_1, ..., \Cli_N$, holding datasets $D_1, ..., D_N$, respectively;
a data buyer $\B$ wanting to know the Shapley value (SV) for each dataset, so that $\B$ can decide how much she should pay for data owners. 
In prior work~\cite{DBLP:journals/pvldb/LiuLL0PS21}, a broker is typically introduced to calculate the SV on behalf of $\B$. However, such a design relies on the honesty of the broker, which could be a strong assumption and challenging to meet realistically. In our framework, we do not assume the existence of a broker.
On the other hand, we assume data owners and the buyer cannot be malicious simultaneously. Specifically, at least all data owners or the buyer are (is) honest. 
Malicious behaviors could occur in both data valuation and payment processes.

Following a setting proposed in~\cite{DBLP:journals/pvldb/JiaDWHGLZSS19}, wherein the buyer has test data to verify the utility of a dataset, we have the same assumption and treat test data as the validation set denoted as $\LL{val}$.
Meanwhile, we mainly focus on evaluating the value of unlabeled data as data sellers may prefer to sell features only, considering that label information is usually more sensitive than features~\cite{DBLP:conf/nips/GhaziGKMZ21}, and labeling data is notoriously costly.
Data owners and the buyer have different goals: each data owner shares her data with the buyer for compensation, and the buyer purchases datasets that are useful for her learning task from data owners. 



Given the above setting and assumptions, we aim to solve the dilemmas of considering both data valuation and fair payment in data marketplaces:
\begin{center}
    {\em The buyer can valuate the data without seeing the data; once the buyer decides to pay, the payment happens atomically. }
\end{center}

\Paragraph{Intuition.} 
Inspired by the work of~\cite{wang2021learnability}, 
we propose to approximate the mapping from a training set to the performance of the ML model trained on the set suing a parametric model, which is referred to as a {\em data utility model} (cf. Section~\ref{sec:data_evaluation}). 
With a well-trained data utility model, the process of calculating utility scores is in fact a prediction process, which can be efficiently executed via a customized MPC protocol (cf. Section~\ref{sec:mpc}). 
The SV can be calculated with the utility scores of different coalitions of data using Equation~\ref{equation: shapley_value}.

Figure~\ref{fig:overview} depicts the {overview} of our scheme considering both data valuation and fair payment.
It roughly works as follows:
\begin{enumerate}
    \item Each $\Cli_i$ pre-shares a small subset of data $d_i \subseteq D_i$ to the buyer $\B$. 
    $\B$ then trains a data utility model $U$ with the data she received from data owners.
    \item Each $\Cli_i$ inputs her data $D_i$ together with a encryption key $k_i$ to MPC. 
    $\B$ inputs (the parameters of) the data utility model $U$ to MPC.
    \item The MPC protocol outputs ($SV_i$, $E(k_i, D_i)$, $H(k_i)$)s to $\B$, where $E()$ is a symmetric-key encryption scheme and $H()$ is a cryptographic hash function. 
    \item $\B$ initiates a hash-locked transaction for each $\Cli_i$, wherein the amount of funds is based on $SV_i$.
    Each $\Cli_i$ claims the funds by putting $k_i$ on-chain.
    \item Each $\Cli_i$ gets the funds, and $\B$ gets $D_i$ by decrypting $E(k_i, D_i)$ with $k_i$.
\end{enumerate}

\section{MPC-Friendly Data Valuation}
\label{sec:data_evaluation}

In this section, we first introduce the notion of data value and explain in detail how to learn an ML performance predictor, which could output model performance corresponding to a training set in an MPC-friendly manner.

\subsection{What is the Notion of Data Value}
In the design of a data marketplace, multiple data owners contribute their data according to the format that buyers require. As different data contribute variously, the main challenge is how to compensate data owners fairly. A natural way is to allocate the revenue based on the usefulness of data from different sources for the buyer's task. Such usefulness is treated as the value of data. There has been a surge of research efforts to formalize the notion of data value~\cite{DBLP:conf/cvpr/Jia0SXDK00S21, DBLP:conf/icml/GhorbaniZ19,DBLP:conf/aistats/Kwon022}, with the two most commonly used being Leave-one-out (LOO)~\cite{DBLP:conf/icml/KohL17} and the Shapley value~\cite{shapely1953value}.

\Paragraph{\textbf{Leave-one-out Method.}}
For $N$ data owners, each of them holding a dataset $D_i, i\in I = \{i,...,N\}$, LOO quantifies the importance of $D_i$ by measure its contribution to the rest of datasets:
\begin{equation}
v_{loo}(D_i)=u(\bigcup_{j\in I} D_j)-u(\bigcup_{j\in I\backslash\{i\}} D_j),
\label{equation: loo_value}
\end{equation}
where $u$ is a function that evaluates the utility of the data. We will formally define $u$ later.

\Paragraph{\textbf{Shapley Value-based Method.}}
The SV characterize the importance of each dataset inspired by cooperative game theory.
A cooperative game is defined by a pair $(I, u)$, where $I=\{1, ..., N\}$ denotes the set of all players and $u:2^N \rightarrow \mathbb{R}$ is the utility function, which maps each possible coalition to a real value indicating the payment of the coalition, i.e., collective payoff a set of players can gain by forming a coalition. 

Given the utility function $u$, the SV of player $i$ is calculated by averaging the marginal contribution of player $i$ to all possible coalitions formed by other players $C \subseteq I\backslash \{i\}$:
\begin{equation}
v_{sv}(u, i)=\frac{1}{N} \sum_{C \subseteq I\backslash \{i\}} \frac{1}{\left(\begin{array}{c}N-1 \\|C|\end{array}\right)}[u(C \cup\{i\})-u(C)] 
\label{equation: shapley_value}
\end{equation}
Suppressing the dependency on $u$, we use $v_{sv}(i)$ to represent the SV of the player $i$.

Transforming the game theory concepts to data valuation, one can think of players as training data sources and the utility function $u(C)$ as a performance metric function, which measure the performance of the model trained on the set of training data $C$. Therefore, the SV of each source measures its contribution to training an ML model. 

We present the formal definition of the data utility function $u$ as below, which is also suitable for the LOO method. Given a learning algorithm $\mathcal{A}$, which takes a set of instances $C$ as input and outputs a classifier $\hat{f}\leftarrow\mathcal{A}(C)$. The metric function $u$ takes $\hat{f}$ as input and outputs the utility of $C$. In the context of machine learning, we often use test accuracy as the metric $u(\hat{f}, \mathcal{V})=\frac{1}{|\mathcal{V}|} \sum_{(x, y) \in \mathcal{V}} \mathbb{1}[\hat{f}(x)=y]$ for a test set $\mathcal{V}$.
With a potential stochastic learning algorithm $\mathcal{A}$ and a corresponding metric function $u$, the data utility function can be defined as $U_{\mathcal{A},u}(C)=\mathbb{E}_{\mathcal{A}}[u(\mathcal{A}(C), \mathcal{V})]$. We omit the subscript when the context is clear and simply write $U(C)$. 

Although LOO has the advantage of low complexity, only $\mathcal{O}\left(N\right)$ where SV has $\mathcal{O}\left(2^{N}\right)$, prior works~\cite{DBLP:conf/aistats/JiaDWHHGLZSS19, DBLP:conf/icml/GhorbaniZ19, DBLP:conf/aistats/Kwon022, DBLP:conf/cvpr/Jia0SXDK00S21} empirically show that SV can better reflect the utility of the data and identify bad and good quality data. Meanwhile, Jia et al.~\cite{DBLP:journals/corr/abs-2205-15466} theoretically prove that SV is more robust against inherent stochasticity of ML models induced by stochastic gradient descent, which produce inconsistent data value rankings across different runs. Furthermore, SV is theoretically proven to satisfy property of rigorous fairness:
(1) if sellers $i$ and $j$ are equivalent in the sense of $v_{sv}(C\cup\{i\})=v_{sv}(C\cup\{j\})$, $\forall C\subseteq I\backslash\{i,j\}$, then $s_{sv}(i)=s_{sv}(j)$, where $s_{sv}(i) (s_{sv}(j))$ is the SV of the player $i (j)$. That means two sellers who are identical in terms of the contribution to ML performance should have the same value. (2) $s_i=0$ if $v_{sv}(C\cup\{i\})=v_{sv}(C)$ for all $C\subseteq I\backslash \{i,j\}$. This means that sellers should receive zero pay-off if their data have no contributions to improving ML performance when combined with all the other data sources.
Those reasons motivate us to adapt SV for data valuation.

\subsection{Efficient Computation of Shapley Value}
\label{ssec:data_utility_sampleing}
Exactly calculating the SV of a set of data requires retraining a model for every coalition of individuals. To solve such a problem, we propose to approximate the data utility function with a parametric model, referred to as the data utility model. 
Next, we explain in detail how we can build a data utility model. It consists of two phases: {\em training dataset construction} and {\em model training}.

\subsubsection{Training dataset construction}
As we described in Section~\ref{sec:overview}, each data owner $\Cli$ pre-shares a small portion of its dataset with the buyer $\B$;
we denote such data collectively as $\LL{tr}$. The training dataset consists of pairs of data subsets and the corresponding utility scores, wherein each subset is sampled from $\LL{tr}$.

We describe the training dataset construction phase in Algorithm~\ref{alg:data_utility_sampling}.
Specifically, given $\mathcal{L}_{tr}$, $\mathcal{L}_{val}$ and a learning algorithm $\mathcal{A}$, $\B$ first randomly samples a subset $S_i \subseteq \mathcal{L}_{tr}$ (line 3) and uses it to train a model $f_i$ with algorithm $\mathcal{A}$ (line 4).
Next, $\B$ evaluates the utility of $f_i$ on $\mathcal{L}_{val}$, denoted as $u(f_i, \mathcal{L}_{val})$. 
Then, the pair of $S_i$ and $u(f_i, \mathcal{L}_{val})$ forms a training instance to learn the data utility model, where $u(f_i, \mathcal{L}_{val})$ is considered the label of $S_i$.
Repeating the above procedure $M$ times, we can get a set $S_{DS}$ to serve as the training dataset for the data utility model.

\renewcommand*{\algorithmcfname}{Algorithm}
\begin{algorithm}[htb]
\caption{Training Dataset Construction}
\label{alg:data_utility_sampling}
\small
\KwIn{training set $\LL{tr}$; validation set $\LL{val}$; learning algorithm $\mathcal{A}$;
} 
\KwOut {utility learning dataset $S_{DS}$ for training data utility model}
\BlankLine
Initialize value dataset $S_{DS}= \emptyset$\\
\For(\hfill \CommentSty{for each subset}){$i=1 \to M$}{
    Sample a subset $S_i\subseteq \LL{tr}$\\
    Train classifier $f_{i} \leftarrow \mathcal{A}(S_i)$\\
    $u(f_{i}, \LL{val}) =\frac{1}{|\mathcal{L}_{val}|} \sum_{(x, y) \in \mathcal{L}_{val}} \mathbb{1}[f_i(x)=y]$\\
    $S_{DS} = S_{DS} \bigcup (S_i, u(\LL{val}))$
}
\Return{$S_{DS}=(X, U)$, where $X=(S_1,...,S_M), U=(u_1,...,u_M)$}
\end{algorithm}



Algorithm~\ref{alg:data_utility_sampling} requires the instances in $\LL{tr}$ to be labeled, which is not the case in our setting.
To resolve this problem, a straightforward method is to label the data before being shared to $\B$. 

\subsubsection{Data Utility Training}
\label{sssec:data_utility_training}

Next, we show how $\B$ trains the data utility model $U$. As each instance in the training dataset is a set, 
we adopt a canonical {\em set function model} - DeepSets~\cite{DBLP:conf/nips/ZaheerKRPSS17} - to train $U$. 
Furthermore, we also need a {\em feature extractor} to extract the feature embedding from each instance in $S_{DS}$, so that the DeepSets model could map the set of feature embeddings to its corresponding utility score.
To this end,  we treat the data utility model $U$ as a composition of a DeepSets model $f_{DS}$ and a feature extractor $G_f$. 

Algorithm~\ref{alg:utility_function_training} details the data utility training phase. We name it \textbf{Labeled Pre-sharing Method}.
Note that different sets $S_i$ and $S_j$, $i\neq j$ may contain the same data points, i.e., $S_i\cap S_j \neq\emptyset$.
In practice, we first fix $G_f$ to optimize $f_{DS}$ (lines 4-5) and then fix $f_{DS}$ to optimize $G_f$ with a single batch/subset (line 6).

\renewcommand*{\algorithmcfname}{Algorithm}
\begin{algorithm}[ht]
\caption{Data Utility Model Training}
\label{alg:utility_function_training}
\small
\KwIn{training set $\mathcal{L}_{tr}$, valuation set $\mathcal{L}_{val}$;
} 
\KwOut {feature extractor $G_f$ and DeepSets model $f_{DS}$}
\BlankLine
Initialize models: feature extractor $G_f$ and DeepSets model $f_{DS}$

$S_{DS}=(X, U)\gets \B$ run Algorithm~\ref{alg:data_utility_sampling} with inputs $\mathcal{L}_{tr}, \mathcal{L}_{val}$

\For(\hfill \CommentSty{for each epoch}){$i=1 \to T$}{
    Fix $G_f$, extract the feature embedding $E_S$ of dataset, $E_S \leftarrow G_f(X)$\\
    Train the DeepSets value model $f_{DS}$ on $(E_S, U)$\\
    Fix $f_{DS}$, optimize $G_{f}$ with $(X, U)$
}
\Return $G_{f}$, $f_{DS}$
\end{algorithm}


\paragraph{\textbf{Data Utility Training with Domain Adaption}}
Recall that we need each $\Cli_i$ to manually label its data before they are shared to $\B$, which requires extra human work. 
To that end, we propose to train the data utility model with {\em domain adaption} (DA) to eliminate the requirement of labels.
The core idea is to train the data utility model on public data that are different from but related to data sources to be purchased and mitigate the impact of domain shift by domain adaption. Adopting domain adaptation is not rare, and such public data exists in many real-world applications. For example, when a service company aims to expand its business into new regions, it is natural to treat data from the original area as public data.


A domain adaption framework usually consists of three components: a feature extractor $Ext$ which maps the inputs to a lower dimension feature embedding, a discriminator $Dis$ which aims to distinguish between source domain and target domain data, and a regressor $Reg$ that takes the output embedding of $Ext$ as input and output predictions. The DA method has two goals: 1) map examples from two different domains to a common feature space; and 2) keep useful information for original tasks. They can be achieved by optimizing the GAN loss $L_{GAN}$ and the prediction loss $L_{DS}$ written as follows:
\begin{equation}
    \begin{aligned}
        \min _{G_{f}} \max _{G_{d}} L_{G A N} =& \sum_{x\in \mathcal{L}_{pub}}\log G_{d}\left(G_{f}(x)\right)\\
        +& \sum_{x\in \mathcal{U}_{tgt}}\log \left(1-G_{d}\left(G_{f}(x)\right)\right),
    \end{aligned}
    \label{equation: GAN_loss}
\end{equation}

\begin{equation}
    \min _{f_{DS}} \min _{G_{f}} L_{DS}
    =\sum_{i=1}^{M}\left\|U_{s}\left(G_{f}\left(S_{i}\right)\right)-u_{i}\right\|^{2}.
\label{equation: LDS_loss}
\end{equation}

To leverage domain adaption to train a data utility model $U=f_{DS} \circ G_f$ useful for evaluating data utility, we treat $G_f$, $f_{DS}$ as $Ext$, $Reg$ separately and introduce a model $G_d$ that serves as a discriminator $Dis$ to help training $G_f$.
The training process of the data utility model $U$ is integrated in the DA training process. 

In more detail, we assume the buyer $\B$ has a public dataset $\mathcal{L}_{pub}$ from the source domain, which can be seen as the training set $\mathcal{L}_{tr}$ to construct $S_{DS}$. Meanwhile, each data owner $\Cli_i$ sends a subset of unlabeled data $d_i\in D_i$ to the buyer. The data consolidation $\mathcal{U}_{tgt} = \{d_1,...,d_N\}$ is treated as samples from the target domain.


The workflow to train the data utility model with domain adaptation is depicted in Algorithm~\ref{alg:utility_function_training_DA}. We name it \textbf{Unlabeled Pre-sharing Method}.
In practice, we train $k$ steps of general domain adaptation training (line 4-6) followed by one step of utility training (line 7-9).
Note that our method can be combined with any state-of-the-art DA frameworks, and we use CyCADA~\cite{DBLP:conf/icml/HoffmanTPZISED18} in this paper. 

\renewcommand*{\algorithmcfname}{Algorithm}
\begin{algorithm}[tb]
\caption{Data Utility Model Training with DA} 
\label{alg:utility_function_training_DA}
\small
\KwIn{public dataset $\mathcal{L}_{pub}$ serves as the training set $\LL{tr}$, valuation set $\mathcal{L}_{val}$ and the unlabeled dataset $\mathcal{U}_{tgt}$ from target domain
} 
\KwOut {feature extractor $G_f$ and DeepSets model $f_{DS}$}
\BlankLine
Initialize models: feature extractor $G_f$, discriminator $G_d$ and DeepSets model $f_{DS}$

$S_{DS}=(X, U)\gets \B$ run Algorithm~\ref{alg:data_utility_sampling} with inputs $\mathcal{L}_{pub}, \mathcal{L}_{val}$

\For(\hfill \CommentSty{for each epoch}){$i=1 \to T$}{
    \For{$k$ steps}{
    Train $G_f$ and $G_d$ with ($\mathcal{L}_{pub}$, $\mathcal{U}_{tgt}$) via optimizing Equation~\ref{equation: GAN_loss}\\
    }
    Fix $G_f$, extract the feature embedding $E_S$ of dataset, $E_s \leftarrow G_f(X)$\\
    Train the DeepSets model $f_{DS}$ on $(E_s, U)$\\
    Fix $f_{DS}$, optimize $G_{f}$ with $(X, U)$ via optimizing Equation~\ref{equation: LDS_loss}
}
\Return $G_{f}$, $f_{DS}$
\end{algorithm}

We discuss how to determine the number of pre-shared data. More pre-shared data leads to training a better utility model but at the same time introduces the potential risk that the buyer gets enough data for free.
To that end, data owners first pre-share a small amount of data, say 1\%, with the buyer. 
The buyer makes public the utility learning dataset $S_{DS}$ as well as the training process of the utility model, such as the initial weight, hyper-parameters, gradients, etc.
The data owner can verify the utility model with public information and re-share more data if she finds that the model generalizes badly on her local data (e.g., the predicted utility score on her local data is significantly different from that on pre-shared data.).
The experiments in Section~\ref{ssec:shared_data} show that 4\% of pre-shared data is enough to train a good utility model, and it is almost impossible for the buyer to learn a good model for her original task using aggregated pre-shared data.

\section{Customized MPC}
\label{sec:mpc}

In this section, we further optimize the MPC circuit design such that the cost of data valuation increases almost unaffected by the number of parties.
We first introduce the background of MPC in Section~\ref{ssec:mpc}. Then, we present in detail how the MPC circuit is being designed in Section~\ref{ssec:optimization_mpc_circuit}. We finally show how to achieve fair payment while ensuring that the data being encrypted is exactly the data being valuated in Section~\ref{ssec:fair_payment}.


\subsection{Background: Multiparty Computation.}
\label{ssec:mpc}
MPC is a type of protocol that allows parties to jointly compute a function $f(x_1,...,x_n) \leftarrow \mathcal{F}(x_1,...,x_n)$ over their input while protecting the privacy of those inputs. It offers the same security guarantee achieved by a trustworthy outside party who receives input from all parties, computes, and returns the corresponding outputs to them so that all parties learn no more except the information that can be inferred from the output and their input. 
Generally speaking, MPC protocols are designed to achieve the following two goals: input privacy and correctness. Input privacy implies that engaging in the protocol does not leak any information about their private data. And correctness comes when the honest parties are guaranteed to compute the correct output or abort if they find an error.

Most MPC protocols make use of the idea of secret sharing, which distribute a secret among a group of parties and each party is allocated a share of that secret. The protocol takes the shares as input and, after a series of calculations, outputs the shares of the result. The result can only be reconstructed when a sufficient number of shares are combined. Two types of secret sharing schemes are commonly used: Shamir secret sharing and additive secret sharing. Shamir secret sharing is also known as $t$-out-of-$n$ secret sharing, which means only when getting the collection of $t$ or more shares can reconstruct the secret, while additive secret sharing needs all shares to reconstruct the secret. MPC protocols have been implemented in several systems with secret sharing schemes, where the most popular one is SPDZ~\cite{DBLP:conf/crypto/DamgardPSZ12}, which is based on additive secret sharing.





\subsection{Private Data Valuation}
\label{ssec:optimization_mpc_circuit}

Running MPC for data valuation has two challenges: (1) the data utility model contains some operations (e.g., comparison) that cannot be supported efficiently by MPC, and (2) the complexity of MPC increases quadratically with the number of parties.
Next we show how we overcome these challenges.


\Paragraph{MPC-friendly operations.}
Recall that the data utility model is a composition of a DeepSets model and a feature extractor. 
The feature extractor contains ReLU and sigmoid, which are slow when running inside the MPC circuit.
Firstly, we use the square function as the activation layer, a common practice to substitute ReLU, which has demonstrated only a slight decrease in accuracy for Neural Network models in~\cite{DBLP:conf/sp/MohasselZ17, 10.1145/3133956.3134056}. 
Notice that that we also use the square function during training to prevent the performance drop caused by the change of activation function. 
Secondly, as the sigmoid function is used in the last layer mapping the output to the values in a range $[0,1]$, we detach it from MPC and run it in plaintext after getting the output of the MPC circuit, which saves much time while having no accuracy loss.

\Paragraph{Optimizations with 2PC.}
We decompose the DeepSets model $f_{DS}$ into three parts $f_{DS}^{trans}, Readout, f_{DS}^{network}$, i.e., 
$$f_{DS} = f_{DS}^{network}(Readout(f_{DS}^{trans})),$$
where $f_{DS}^{trans}$ and $f_{DS}^{network}$ are two fully connected layers with different parameters, and Readout is a function to aggregate the representations. $U$'s forward propagation process can be described as follows:
\begin{itemize}
    \item Encode each instance $x_m \in C$ into a low-dimension feature embedding $G_f(x_m)$;
    \item Transform each embedding $G_f(x_m)$ into some representations $\varphi(x_m)$ by $f_{DS}^{trans}$, i.e., $\varphi(x_m) = f_{DS}^{trans}(G_f(x_m))$;
    \item Aggregate the representations of all embeddings and input the result into $f_{DS}^{network}$ to get the utility score.
\end{itemize}
Notice that the feature extractor $G_f$ and $f_{DS}^{trans}$ only involve one data owner. 
Thus they can be done via 2PC circuits. 
Although the remaining process ($f_{DS}^{network}$) still requires an MPC circuit, it consists of only a few fully connected layers, which is more efficient to implement compared with convolution layers in $G_f$ and $f_{DS}^{trans}$ for image data.
Meanwhile, note that the $Readout$ function is a simple permutation-invariant function such as summation or average. We use average here.

To scale to more data owners, we run most of the operations in 2PC instead of MPC.
To this end, we transfer a big portion of the MPC circuit to 2PC circuits while keeping the same level of security.
The core idea is to share the secret shares of the output of the 2PC protocol with $m$ ($m>2$) data owners and treat them as input to the MPC protocol.
We denote a shared variable $x$ as $\xx$ and all shares calculated via $n$ secret sharing scheme are referred as $\xx_n$. The individual share of $\xx_n$ held by $\Cli_i$ as $\xx_{i/n}$. 
The details of the conversion from $\xx_2$ to $\xx_n$ are shown in Protocol~\ref{alg:2pc2mpc}. 
For simplicity, we present the conversion of $\xx_2$ held by $\Cli_i$ and $\Cli_j$ to $\xx_n$ held among data owners $1,...,n$, where $i \neq j$ and $1\leq i, j\leq n$. 
$\Cli_i$ and $\Cli_j$ first share the shares of $\xx_{1/2}$ and $\xx_{2/2}$ with other data owners (line 1-2), who can directly add the shares they received to obtain the individual $\xx_{k/n}, k\in\{1,...,n\}$.

\renewcommand*{\algorithmcfname}{Protocol}
\begin{algorithm}[tb]
\caption{Conversion of 2 shares calculated via 2 secret sharing scheme to $n$ shares}
\label{alg:2pc2mpc}
\small
\KwIn{$\Cli_i$ holds $\xx_{1/2}$, $\Cli_j$ holds $\xx_{2/2}$, a $n$ secret share scheme $\mathcal{S}_n$} 
\KwOut {$n$ shares $\xx_{n}$ of $x$ held by $\Cli_i,..., \Cli_n$, separately, where $1\leq i, j\leq n$}
\BlankLine
$\Cli_i$ runs $\{s_1^0,...,s_n^0\} \leftarrow \mathcal{S}_n(\xx_{1/2})$,  and send $\{s_1^0,...,s_{i-1}^0, s_{i+1}^0,...,s_n^0\}$ to other players 

$\Cli_j$ runs $\{s_1^1,...,s_n^1\} \leftarrow \mathcal{S}_n(\xx_{2/2})$,  and send $\{s_1^1,...,s_{j-1}^1, s_{j+1}^1,...,s_n^1\}$ to other players

\For{$k \in \{1,...,n\}$}{
$\Cli_k$ run $\xx_{k/n} = s_k^0 + s_k^1$ to get the individual share 
}
\end{algorithm}

In practice, we adopt a widely-used system SPD$\mathbb{Z}_{2^k}$~\cite{10.1007/978-3-319-96881-0_26} to implement our protocol, which build MPC protocol with additive secret sharing. Thus, only collecting all shares can reconstruct the secret $x$.

\subsection{Fair Payment}
\label{ssec:fair_payment}
The fair payment is achieved using the technique of \textit{hash-locked} transactions.

The hash-locked transaction~\cite{Hashlock} is build upon the Bitcoin system. A hash-locked transaction can be redeemed only after the redeemer provides the preimage of a given hash.
Specifically, to create such a transaction, the payer attaches the value of the hash $h$, such that $h=H(m)$ (where $H$ is a cryptographic hash function) for some $m$, and a script that specifically asks for the preimage of $h$  to the transaction. And the transaction can only be finalized by providing a preimage $m$, which hash is exactly $h$. Meanwhile, combined with the technique of time locks, it guarantees that the funds return to the payer if not redeemed after the deadline, creating bidirectional micro-payment channels between the payer and the receiver~\cite{DBLP:journals/ipl/Delgado-SeguraP18}. The use of hash time-locked contracts allows to securely transfer Bitcoins between parities while minimizing the transactions stored on the blockchain, gaining a lot of popularity in the Bitcoin system.


Adopting the hash-locked transactions in our setting, each data owner $\Cli_i$ are required to calculate $E(k_i,D_i)$ and $H(k_i)$, where $E()$ is a symmetric-key encryption scheme, $k_i$ is an encryption key and $H()$ is a cryptographic hash function. If the buyer is willing to purchase data from $\Cli_i$, she issues a $\textit{hash-locked}$ transaction~\cite{Hashlock} on the blockchain, which can only be redeemed by providing the preimage of $H(k_i)$. 
$\Cli_i$ then post $k_i$ to the blockchain to redeem the payment. 
After getting $k_i$, the buyer can decrypt the encryption to obtain the purchased data.

ZKCP uses \textit{zero-knowledge proof} (ZKP) to guarantee the correctness of the data and the encryption key. Specifically, each $\Cli_i$ is required to provide the following ZKP:
\begin{itemize}
    \item $E(k_i,D_i)$ is the encryption of the data that the buyer intents to buy; 
    \item $H(k_i)$ is the hash of encryption key that can be used to decrypt $E(k_i,D_i)$.
\end{itemize}
However, it does not guarantee that the encrypted data is exactly the evaluated data.

In our work, the ``proof'' is inherent inside the MPC circuit.
Specifically, treat the MPC circuit as a black box, which takes data $D$, the encryption key $k$, as well as the data utility model $U$ as inputs and outputs $E(k,D)$, $H(k)$ together with $SV$s.
In other words, $E(k,D)$ and $H(k)$ are obtained collaboratively by each data owner and the buyer via MPC. 
As the MPC protocol is secure against malicious players, it guarantees that no one can modify the output~\cite{10.1007/978-3-319-96881-0_26}.
Thus, it is guaranteed that the data being encrypted is exactly the data being evaluated, and the encryption key is exactly the preimage of the hash value.

\section{Experiments}
\label{sec:experiments_for_utility}
In this section, we conduct experiments to evaluate the effectiveness, scalability, and robustness of our approach to data valuation. We also evaluate the efficiency for the MPC circuit by simulating the scenario of domestic transactions, where data owners and the buyer in the same country, and cross-border transactions (e.g., China and America), where data owners and the buyer in the different countries.

\subsection{Experimental Setup}
We conduct experiments on 5 datasets 
covering census records: \textsl{a9a}~\cite{zeng2008fast} and images: \textsl{MNIST}\cite{lecun-mnisthandwrittendigit-2010}, \textsl{USPS}~\cite{Dua:2019}, \textsl{CIFAR-10}~\cite{krizhevsky2009learning}, \textsl{STL-10}~\cite{coates2011analysis} as a benchmark. 

\Paragraph{Experiments for Data valuation}
All experiments are implemented at least 3 times on a single server equipped with a 2.5 GHz CPU and Tesla V100 GPU.
For the sake of efficiency, we implement the prediction process of the data utility model in plaintext. Meanwhile, when experiments are simulated on more than 8 data owners, we approximate SV using a sampling-based approach that requires $O(NlogN)$ samples to achieve the desired approximation error~\cite{DBLP:journals/corr/MalekiTHRR13}. We adapt the Logistic Regression algorithm as the proxy algorithm to construct the training dataset (cf. Algorithm~\ref{alg:data_utility_sampling}) for efficiency.

\Paragraph{Experiments for MPC circuit}
Each participant is simulated as a single server equipped with a 2.5 GHz Intel CPU and 32 GB RAM. The bandwidth is set to 100 Mbps.
We implement the MPC protocol with the MP-SPDZ library~\cite{mp-spdz}, which includes more than 30 variants of the MPC protocol with the same Python-based high-level programming interface. Specifically, we implement with the SPD$\mathbb{Z}_{2^k}$ protocol~\cite{10.1007/978-3-319-96881-0_26}.
The data encryption and hash are implemented with the Bristol Fashion circuits~\cite{BristolFashion}, which have also been implemented well in the MP-SPDZ library.


\begin{table}[htb]
\begin{tabularx}{8.5cm}{ccc}  
\toprule 
\textbf{Dataset} & \textbf{\makecell{Number of Data Owners}} & \textbf{\makecell{Size of Group}}
\\
\midrule  
\ a9a &  40 & 400\\ 
\ MNIST &  60 & 300\\ 
\ USPS &  50 & 100\\ 
\ CIFAR-10 &  40 & 400\\ 
\bottomrule 
\end{tabularx}

\begin{tabularx}{8.5cm}{ccccc}  
\toprule 
\textbf{\makecell{Source\\Domain}} &\textbf{\makecell{Target\\Domain}} & $|\LL{pub}|$ & \textbf{\makecell{Number of\\ Data Owners}} & \textbf{\makecell{Size of \\ Group}}
\\
\midrule 
\ MNIST &USPS &\numprint{7000} &  50 & 100\\ 
\ STL-10 & CIFAR-10 & \numprint{4000}& 40 & 400\\ 
\bottomrule 
\end{tabularx}   
\caption{Data division for the labeled pre-sharing method (Up) and the unlabeled pre-sharing method (Low).}
\label{tab:dataset_setting}
\end{table}  


\subsection{Effectiveness of data valuation algorithms}
\label{sec:Effectiveness}

We start by exploring the effectiveness of our data valuation algorithms.
We bootstrap dataset to synthesize data owned by data owners and the buyer. For simplicity, we call the data owned by each data owner as a \textbf{group}. Each group can be seen as a data owner who wants to sell to the broker. Unless otherwise indicated, each data owner pre-shares $10\%$ data with the buyer to train the data utility model. We also randomly selected 300 samples from the remaining training dataset for the target domain as a validation set $\mathcal{L}_{val}$. The experimental setting for data division is shown in Table~\ref{tab:dataset_setting}.


During the experiments, we manually add different noises to datasets owned by each data owner. Specifically, for the \textsl{a9a} dataset, which is a tabular dataset and all features are binary variables, we assign to each group a probability $p\in [0,1]$ such that it has the probability $1-p$ to flip the feature values. For the data owner $\Cli_i, i\in\{1,...,N\}$, $p_i = \frac{i-1}{N}$. We add Gaussian noise with scale $\sigma_i = 1 + 9 \cdot\frac{i}{N}$ for each $\Cli_i$ for the other digital datasets. 

\begin{wrapfigure}{l}{0.22\textwidth}
\centering
\includegraphics[width=0.2\textwidth]{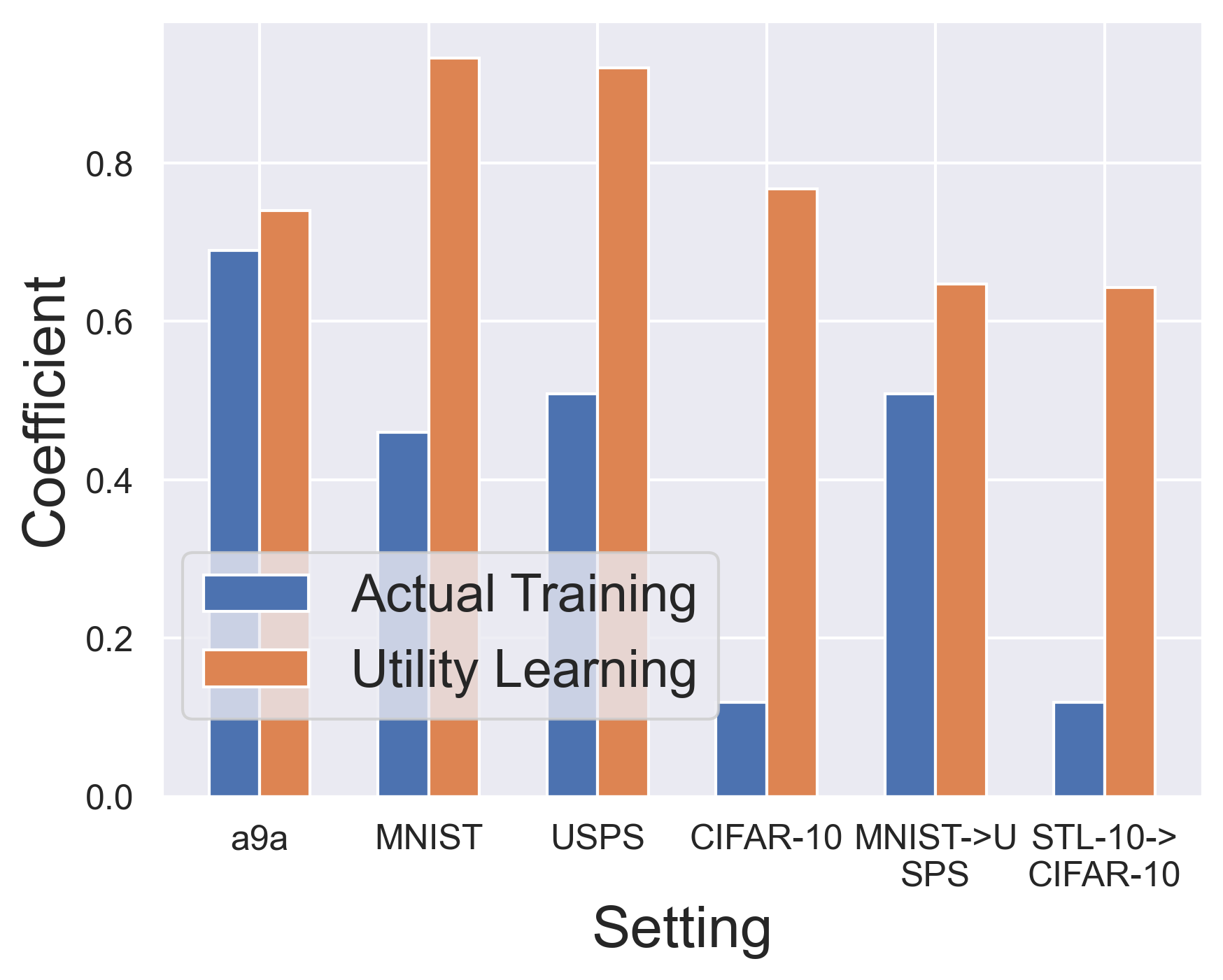}
\caption{Correlation coefficient of the rank of noise magnitude and the rank of SV.} 
\label{fig:coefficient}
\end{wrapfigure}

We show the effectiveness by calculating the correlation coefficient~\cite{Spearman} of the rank of noise magnitude and the rank of SV. The results are presented in Figure~\ref{fig:coefficient}. It shows that SV calculated in our method has a coefficient bigger than $0.5$ in all settings. Although the data quality is not strictly related to added noise, the data with more noise tend to be low quality. Hence, a high coefficient implies that our method effectively evaluates data quality. 

\begin{figure*}[!htbp]
\centering
\setlength\tabcolsep{1.0pt}
\begin{tabular}{ccccc}
    & a9a & MNIST & USPS & CIFAR-10\\[-0.5ex]
    \rotatebox[origin=c]{90}{\parbox{2.5cm}{Removing low\\ value data}}&
    \subfloat{
        \includegraphics[width=0.22\textwidth,valign=c]{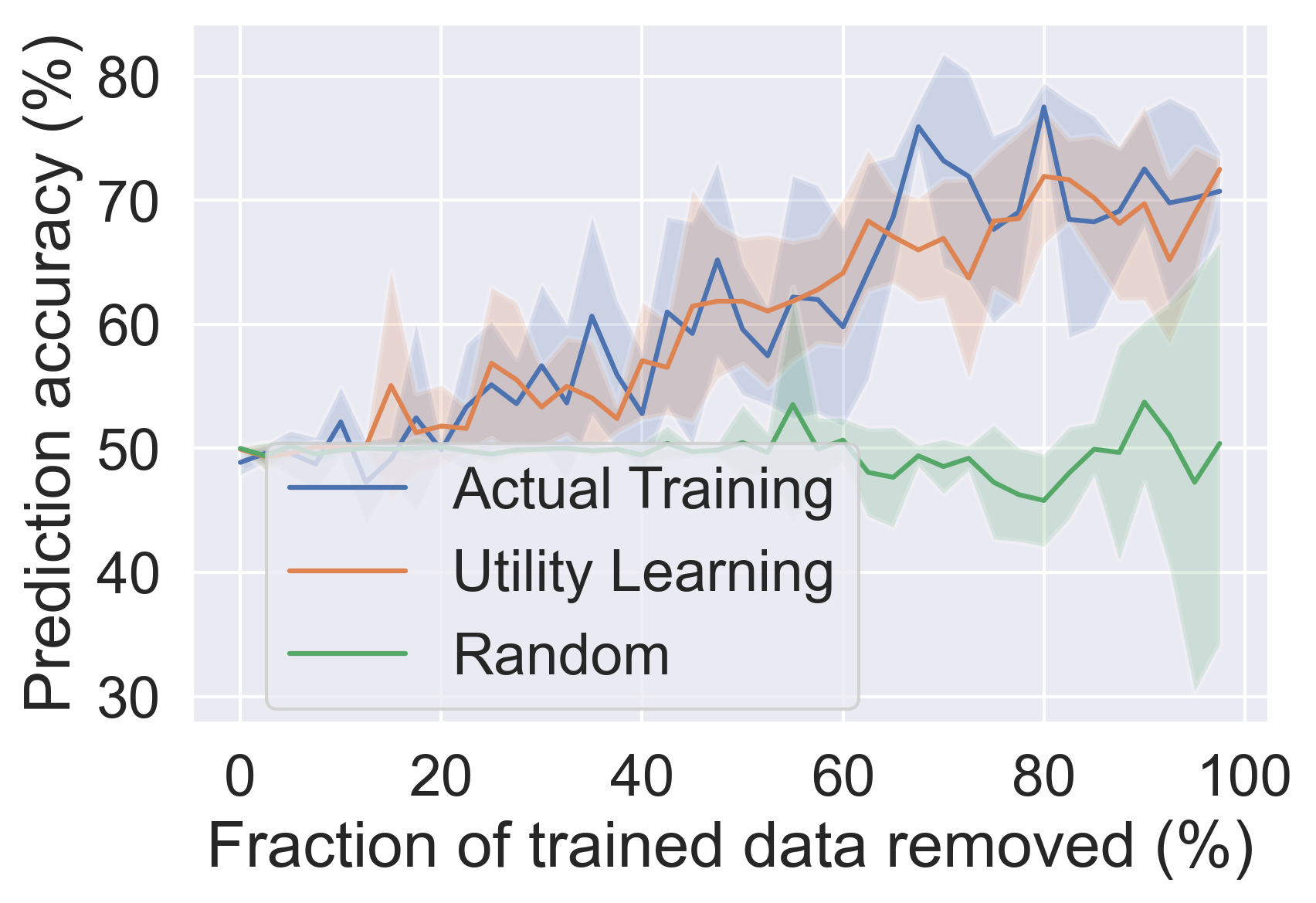}} &
    \subfloat{
        \includegraphics[width=0.22\textwidth,valign=c]{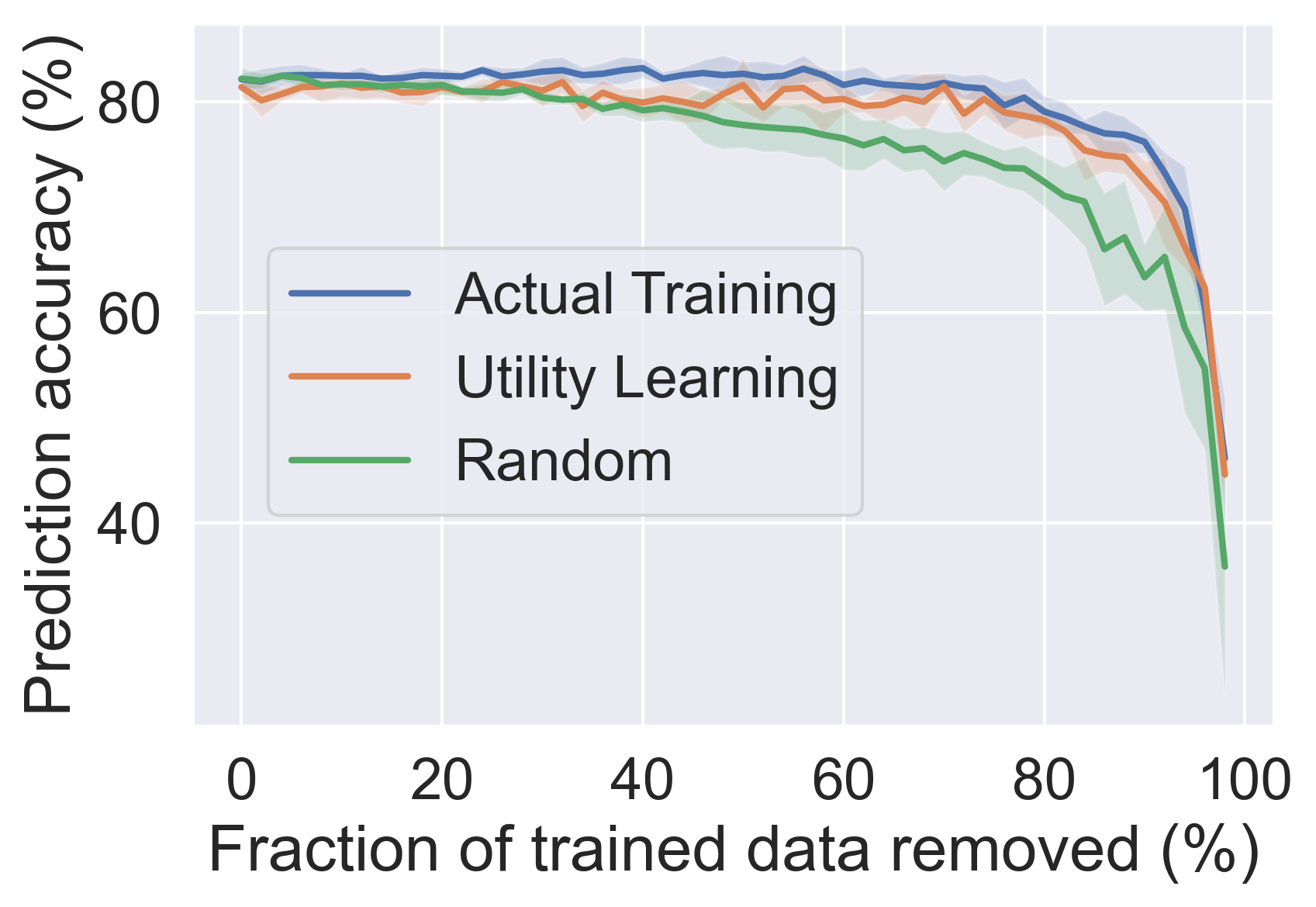}} &
    \subfloat{
        \includegraphics[width=0.22\textwidth,valign=c]{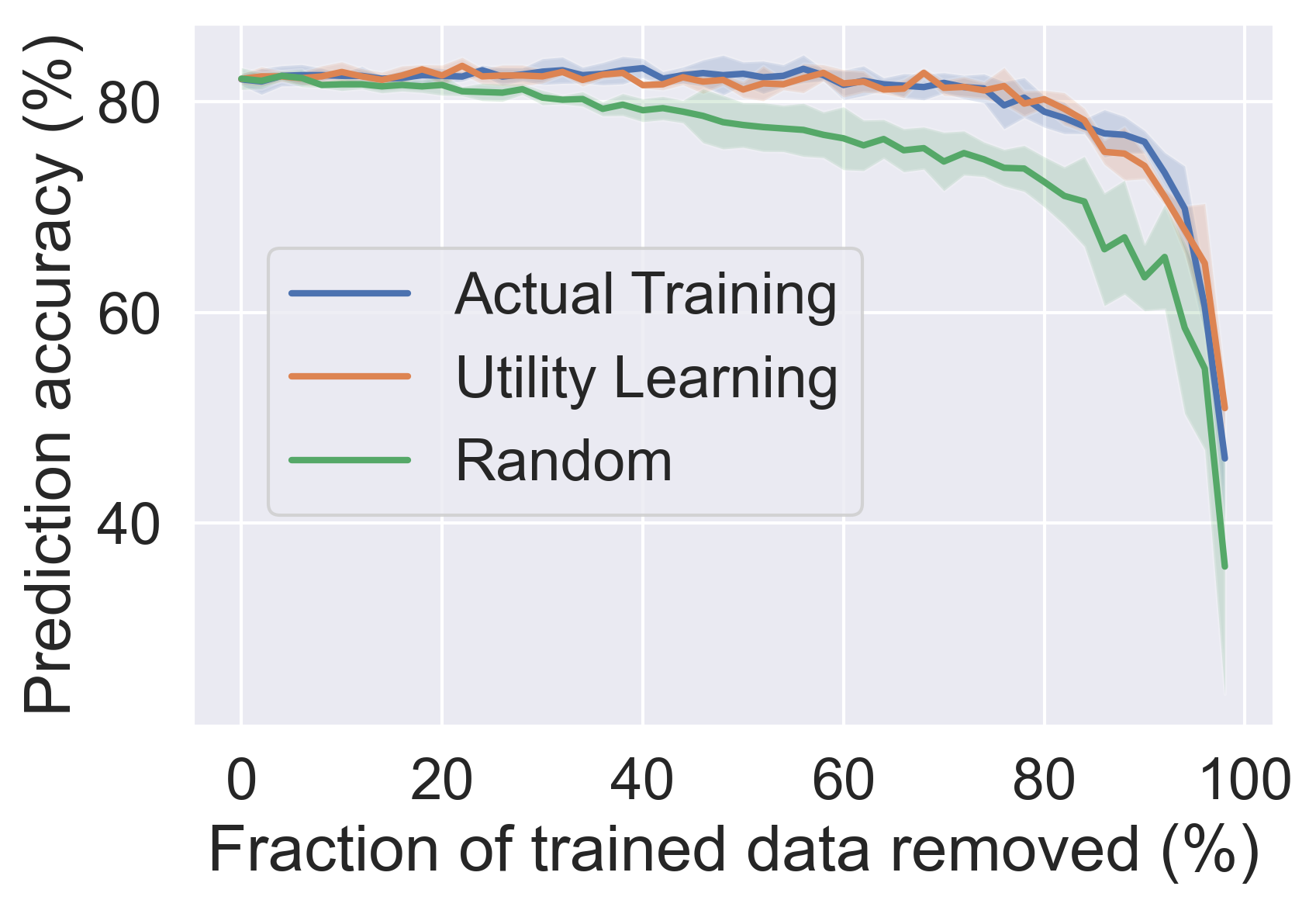}}&
    \subfloat{
        \includegraphics[width=0.22\textwidth,valign=c]{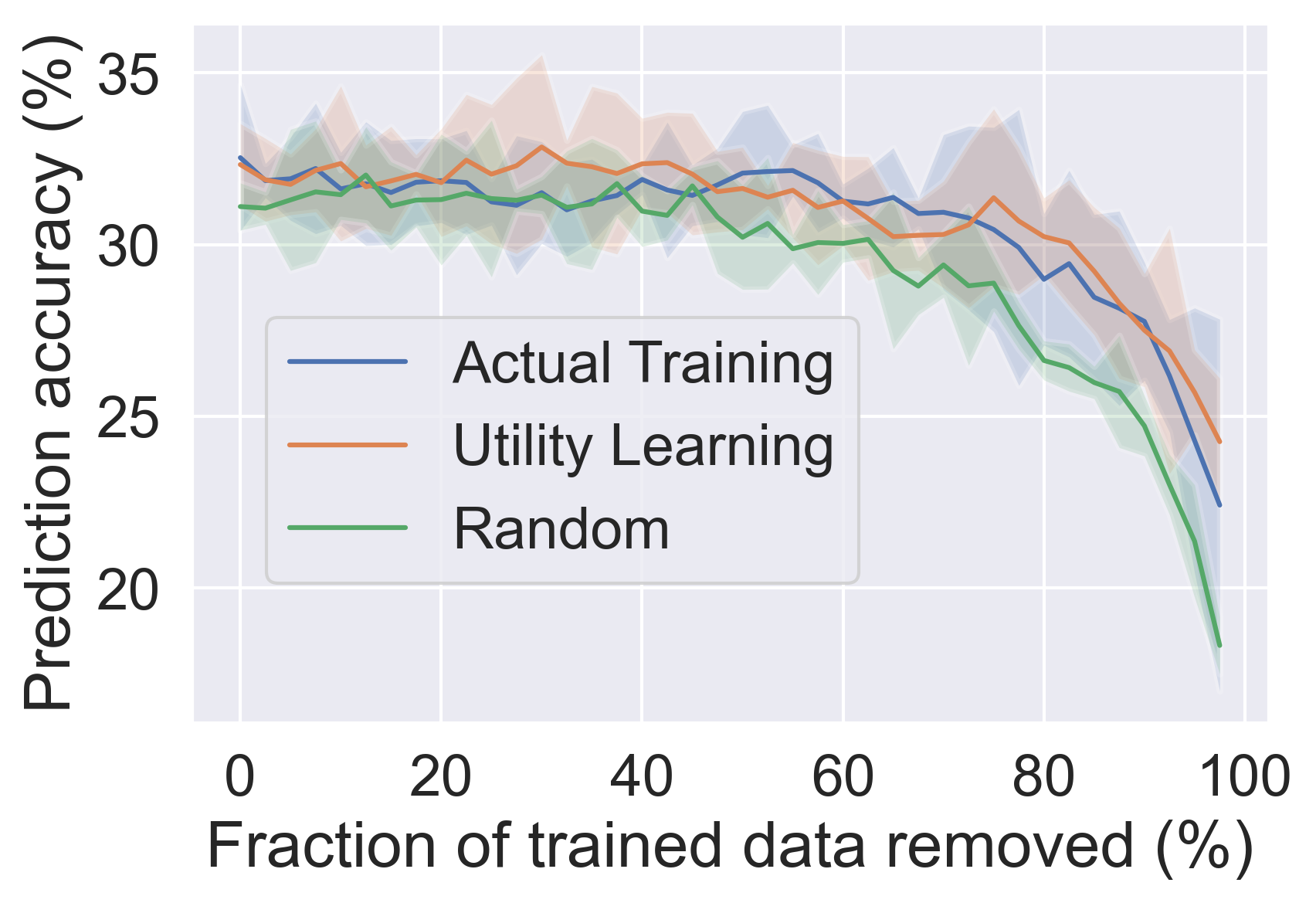}}\\
\addlinespace
    \rotatebox[origin=c]{90}{\parbox{2.5cm}{Removing high\\ value data}}&
    \subfloat{
        \includegraphics[width=0.22\textwidth,valign=c]{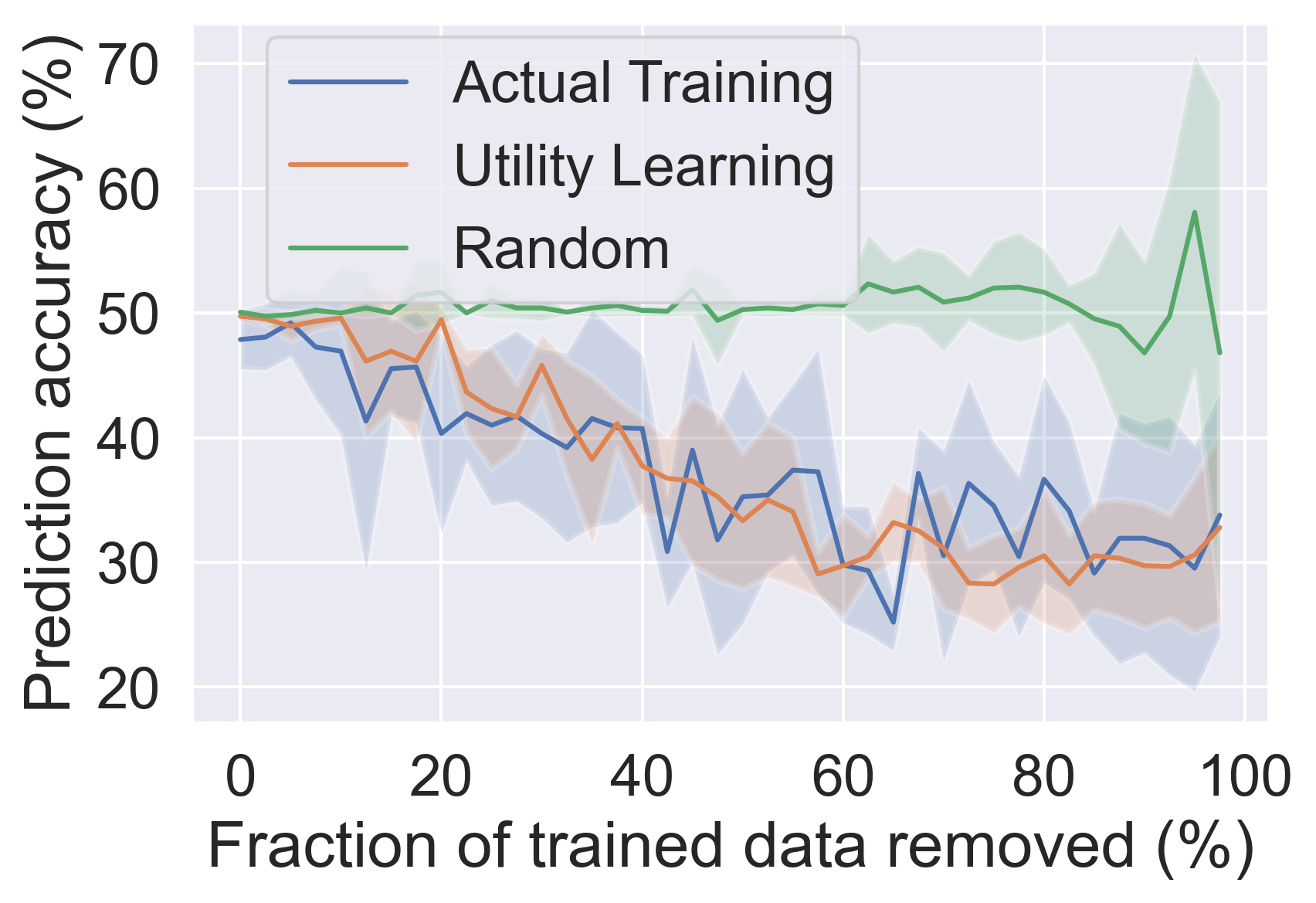}} &
    \subfloat{
        \includegraphics[width=0.22\textwidth,valign=c]{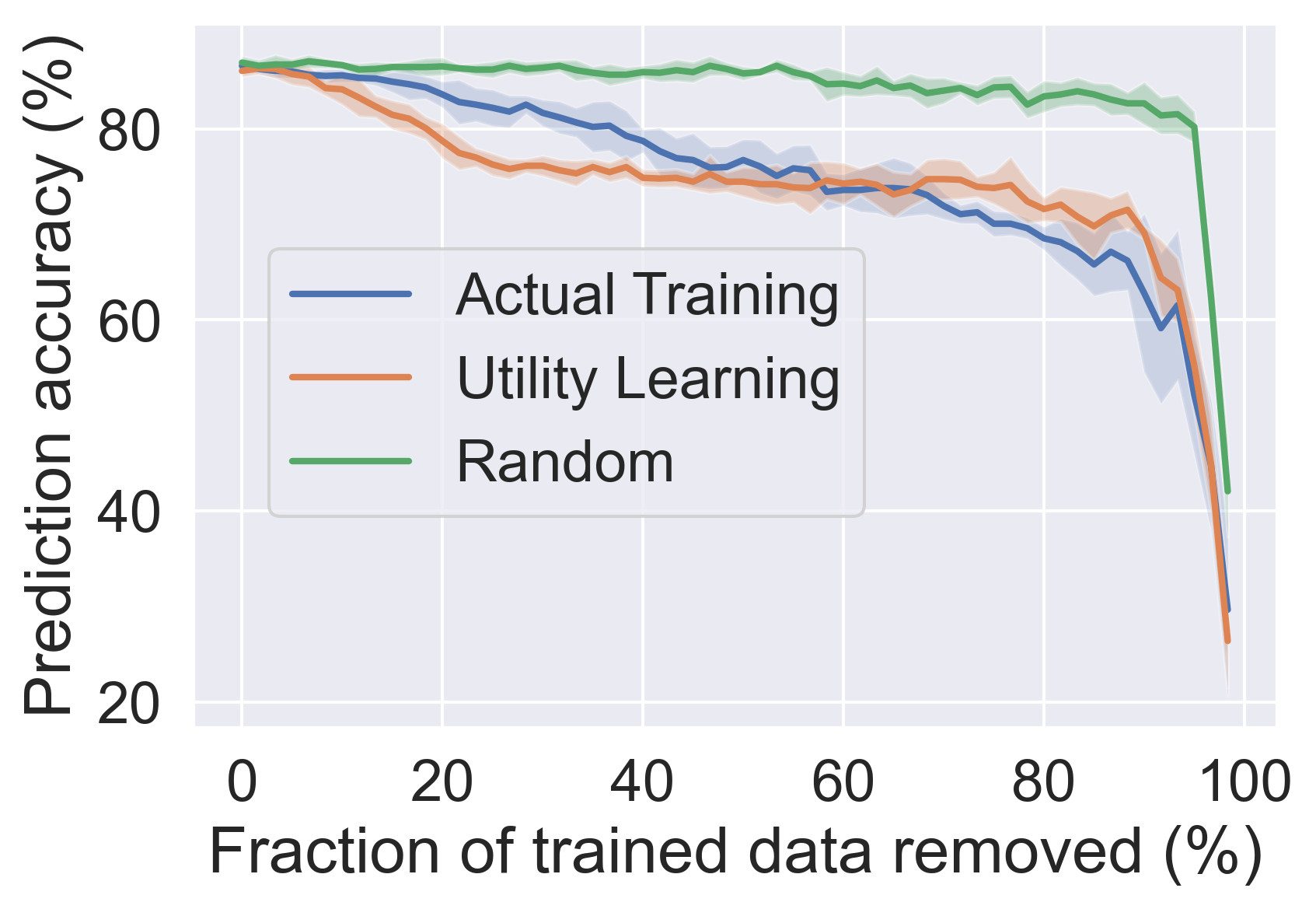}} &
    \subfloat{
        \includegraphics[width=0.22\textwidth,valign=c]{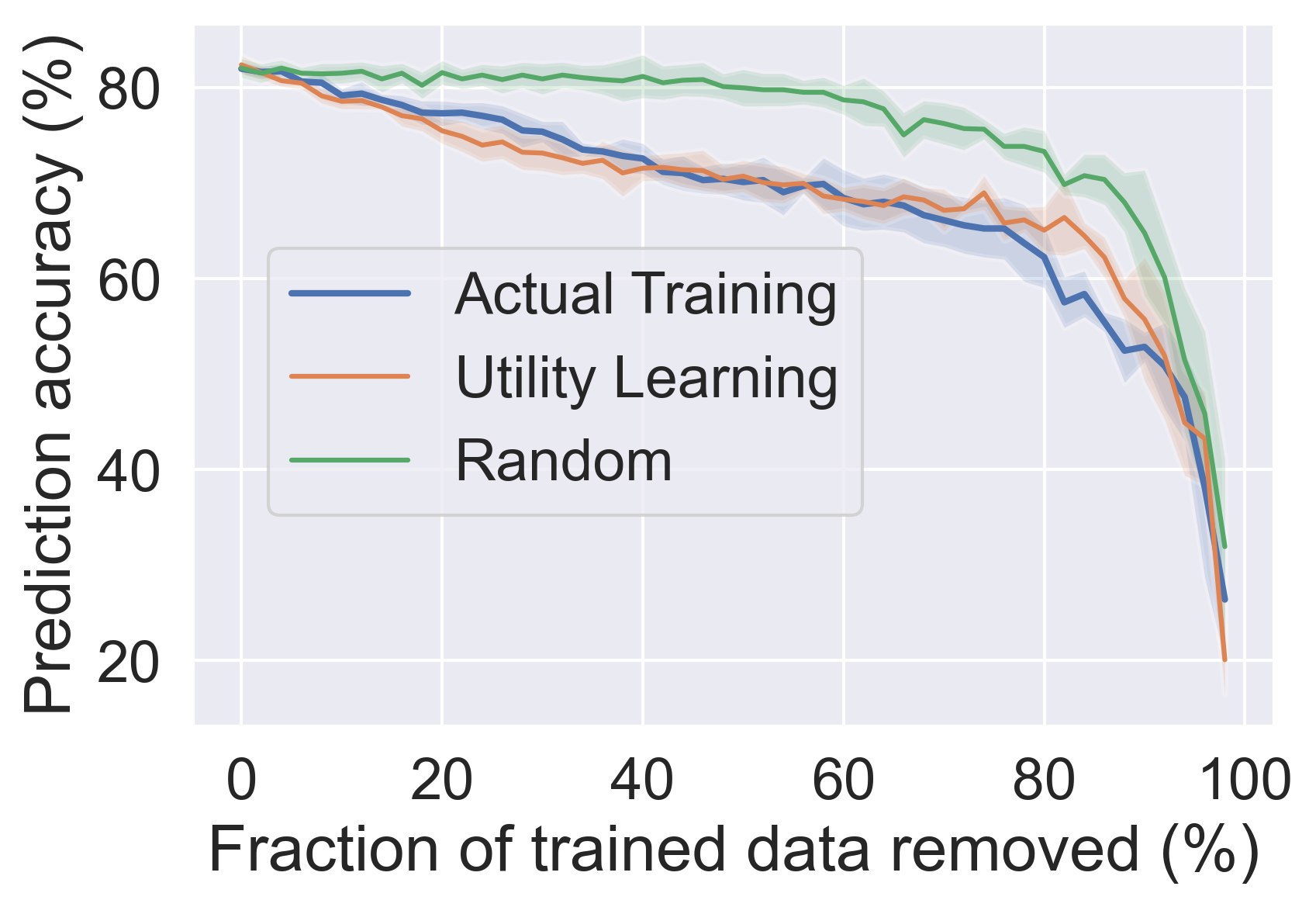}} &
    \subfloat{
        \includegraphics[width=0.22\textwidth,valign=c]{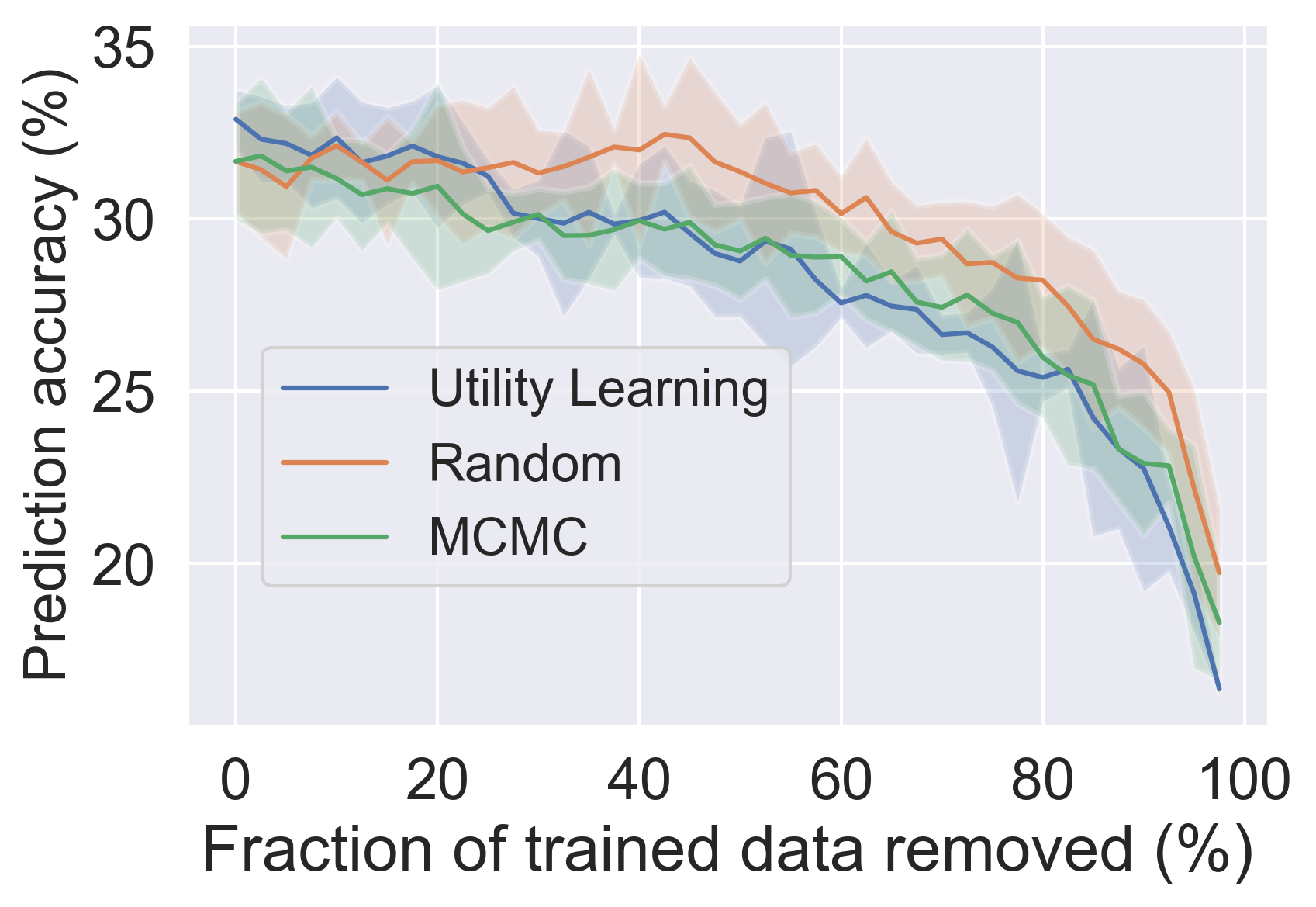}}
\end{tabular}
\caption{Experiments for labeled pre-sharing method. Removing data in ascending order (Up) or descending (Low) order of SV.}
\label{fig:Effectiveness_labeled}
\end{figure*}

\begin{figure}[!htbp]
\centering
\setlength\tabcolsep{1.0pt}
\begin{tabular}{ccc}
    & MNIST$\Rightarrow$USPS & STL-10$\Rightarrow$CIFAR-10\\[-0.5ex]
    \rotatebox[origin=c]{90}{\parbox{2.5cm}{Removing low\\ value data}}&
    \subfloat{
        \includegraphics[width=0.21\textwidth,valign=c]{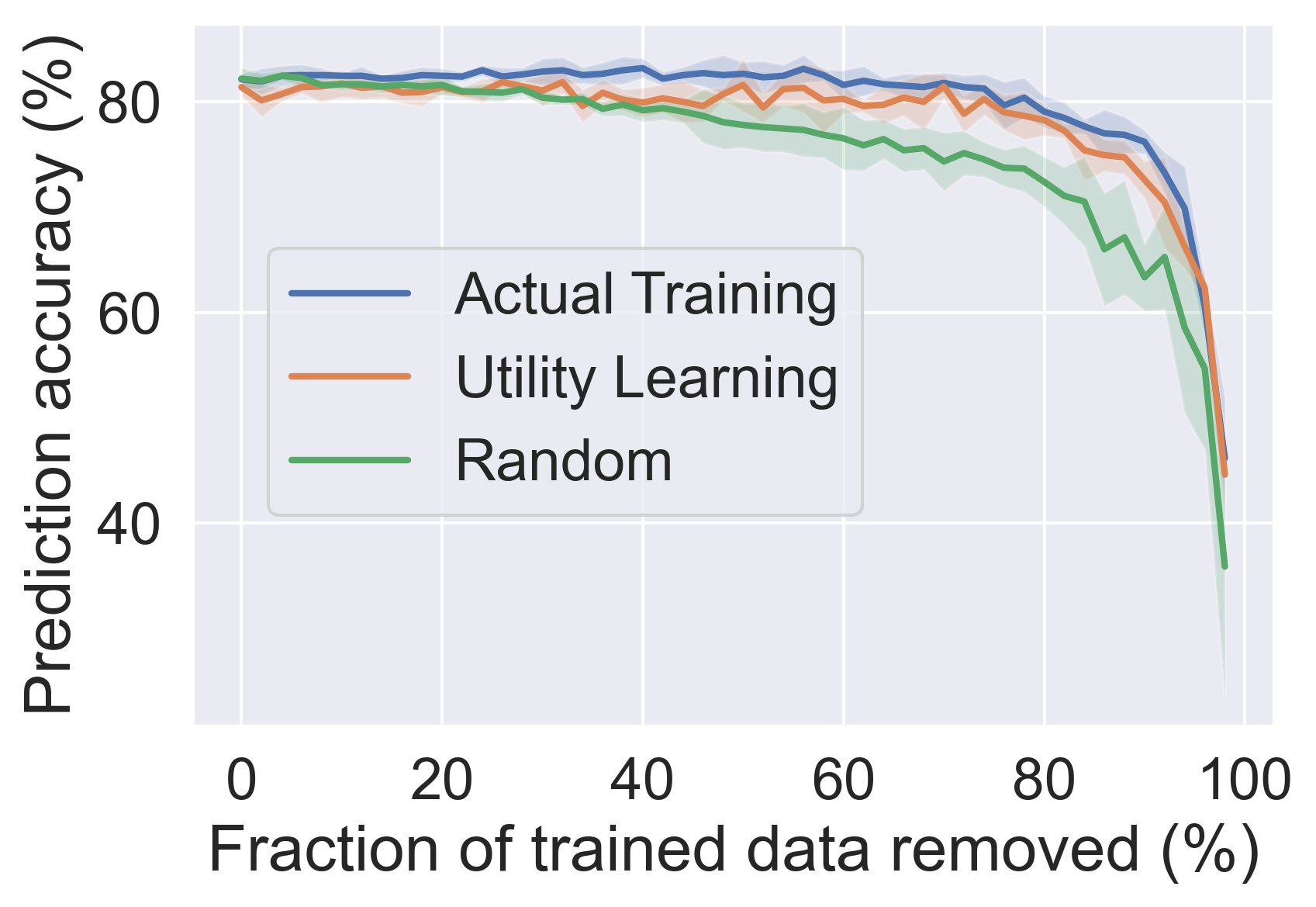}} &
    \subfloat{
        \includegraphics[width=0.21\textwidth,valign=c]{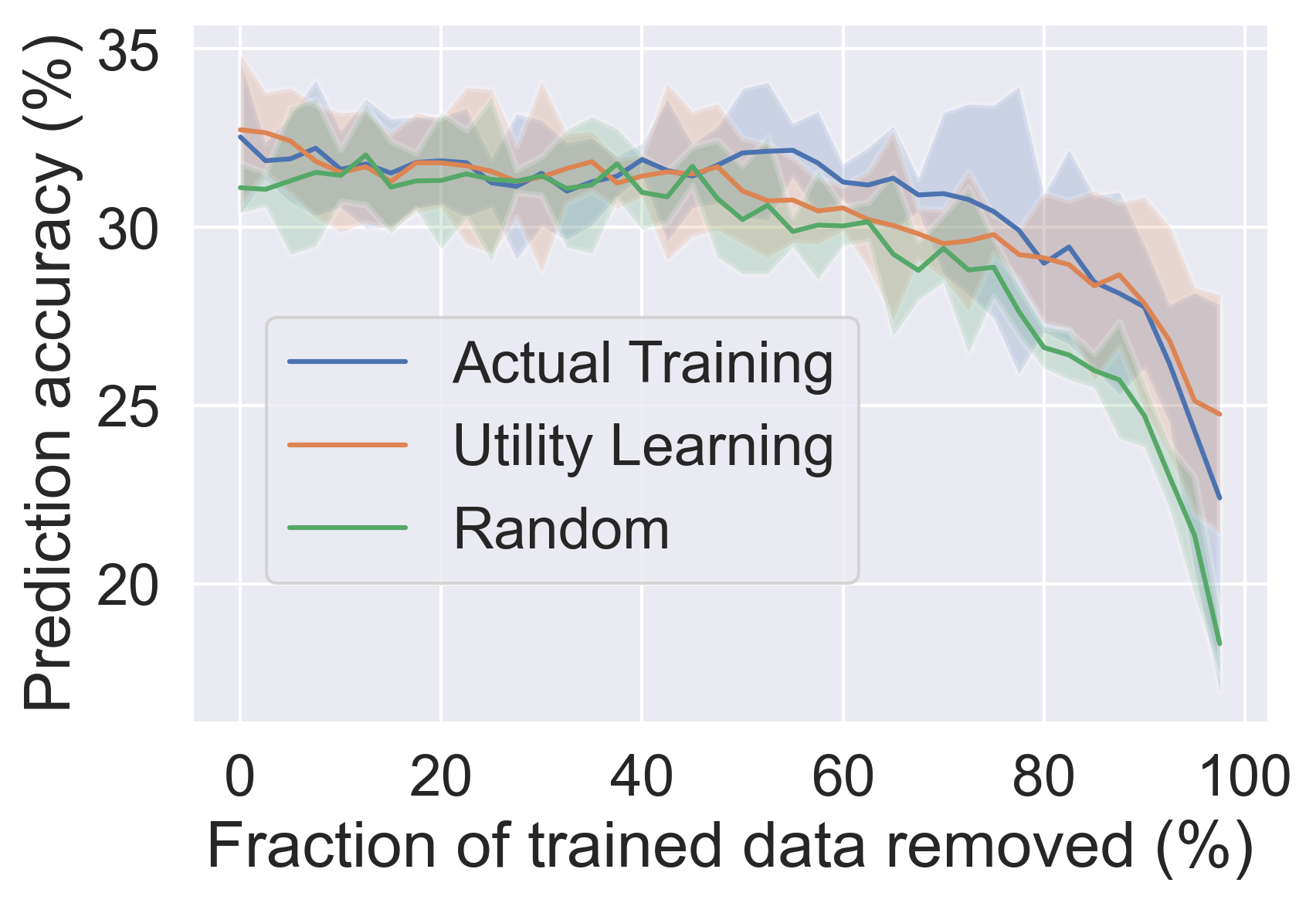}} \\
\addlinespace
    \rotatebox[origin=c]{90}{\parbox{2.5cm}{Removing high\\ value data}}&
    \subfloat{
        \includegraphics[width=0.21\textwidth,valign=c]{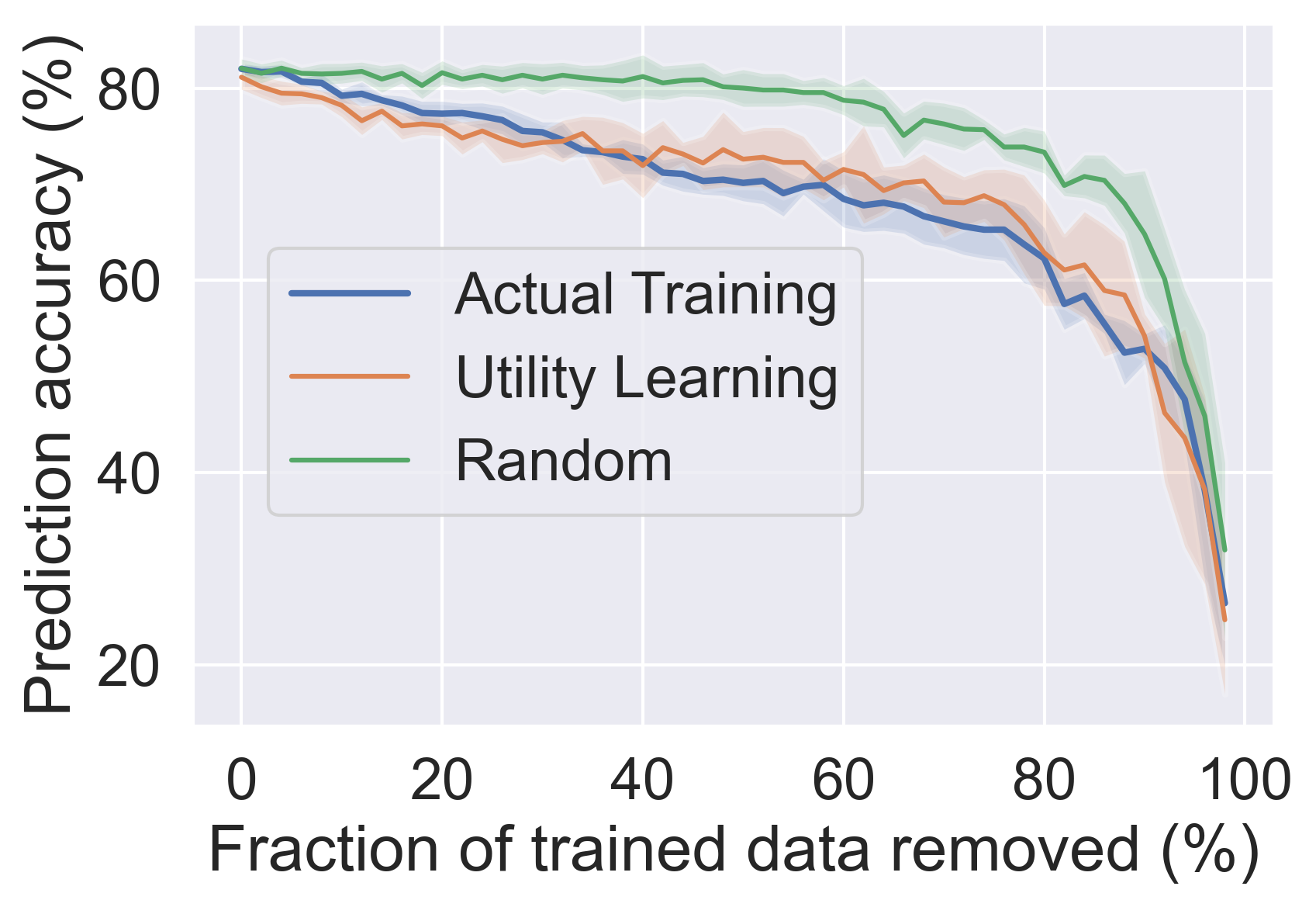}} &
    \subfloat{
        \includegraphics[width=0.21\textwidth,valign=c]{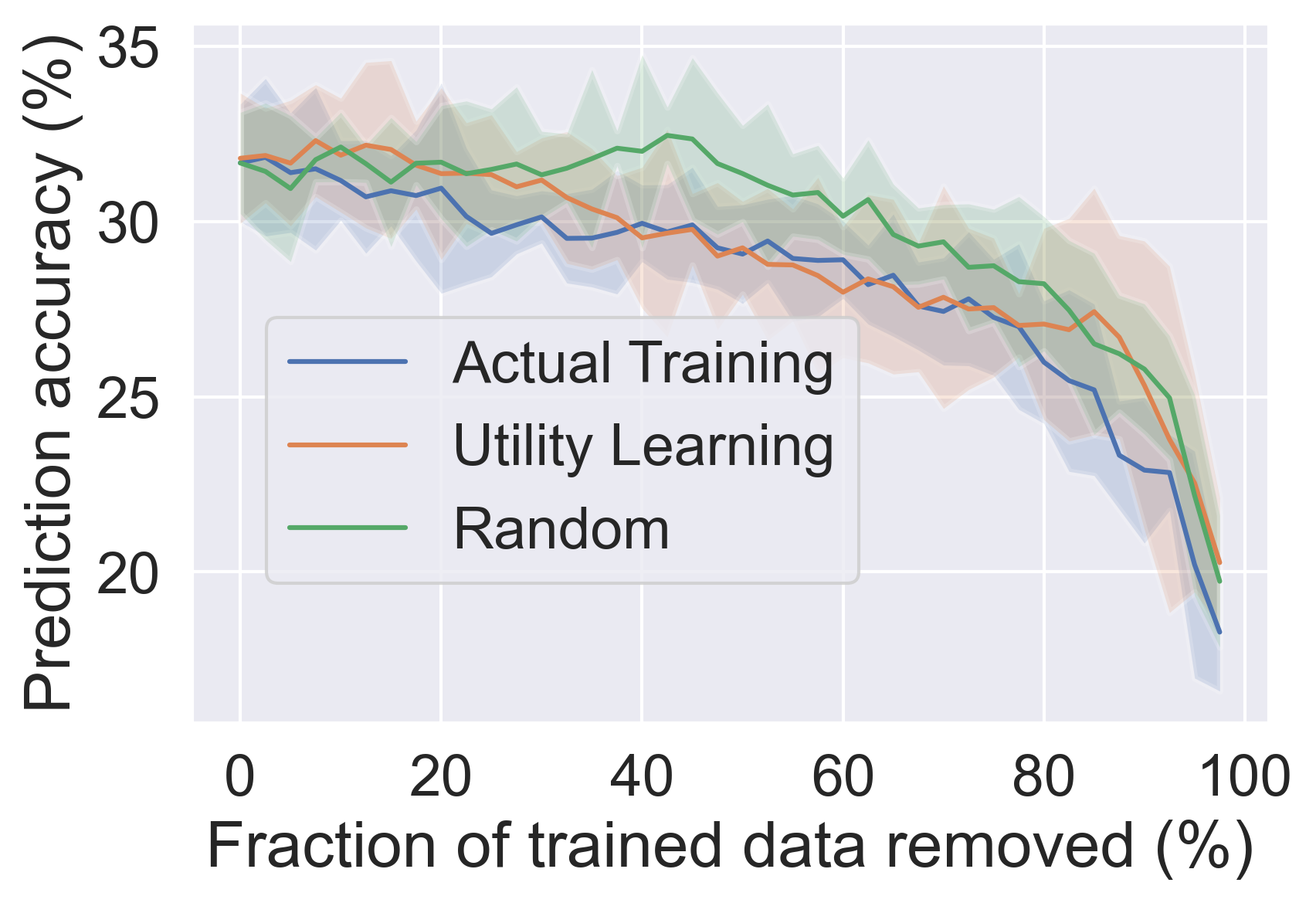}}
\end{tabular}
\caption{Experiments for the unlabeled pre-sharing method. Data removal in ascending order (Up) or descending order (Low) of SV.}
\label{fig:Effectiveness_unlabeled}
\end{figure}

Next, we conduct experiments of removing data in the same group until only one group remains according to the SV calculated in advance and observe the test accuracy in $\LL{val}$ of the model trained with the remaining data. Intuitively, a better SV estimate can better identify the importance of each group. Therefore, when the data with the highest (lowest) SV estimates are removed, a better data value estimation method would lead to a faster (slower) performance drop. The results are depicted in Figure~\ref{fig:Effectiveness_labeled} and Figure~\ref{fig:Effectiveness_unlabeled}.

It shows that our methods perform well in identifying the quality of data groups and have similar performance to calculating SV via actual training.
Notice that the accuracy of the model trained on the a9a dataset increases with the removal of low-SV data, probably because we add too much noise to the data, making it difficult to train an effective model. 



\subsection{Scalability of data valuation algorithms}

Next, we vary the number of data owners and the size of each group to verify the scalability of our methods in data removal tasks. 
For simplicity of presentation, we propose an intuitive metric named \textsl{effectiveness score} to represent the effectiveness of our methods. The metric is calculated as follows:
\begin{equation}
    a = 
    \begin{cases}
        \sum\limits_{t=1}^{T}(acu_l^t-acu_r^t)\cdot\frac{1}{T}~ \mathit{if}~\text{Removing low-value data},\\
        \sum\limits_{t=1}^{T}(acu_r^t-acu_h^t)\cdot\frac{1}{T}~ \mathit{if}~\text{Removing high-value data},
    \end{cases}
    \label{equation:effectiveness_score}
\end{equation}
where $T$ is the number of removal times, $acu$ represents accuracy, the subscripts $l, r, h$ represent the removal of low SV data, the random removal and the removal of high SV data.
Intuitively, $a$ can be treated as the area between two accuracy lines caused by data removing according to SV and randomly removing.
A positive $a$ represents that our data valuation method is effective.
The results are presented in Table~\ref{tab:scalability}.  All of them show that our methods are effective even with different numbers of data owners and different sizes of groups, explaining their good scalability.

\subsection{Robustness of data valuation algorithms}
We conduct a series of experiments to verify the robustness of our methods. Specifically, given some data owners, who own imbalance datasets or adversarial datasets, we experimentally prove that our data valuation methods still have a good performance. We further consider two types of malicious data owners: 1) data owners who intentionally select good data to pre-share and 2) data owners who share incorrect labels to train the data utility model for the labeled pre-sharing method. A malicious data owner may pre-share "poisoned" data during the sample step, specifically designed to allow the data owner to lower the value of other data owners' data. However, developing such an attack is beyond the scope of this paper, and we leave it for future work.

\begin{figure}[htbp]
\centering
\setlength\tabcolsep{1.0pt}
\begin{tabular}{ccc}
    & CIFAR-10 & \makecell{ STL-10 $\Rightarrow$ CIFAR-10}\\[-0.5ex]
    \rotatebox[origin=c]{90}{\parbox{2.5cm}{Removing low\\ value data}}&
    \subfloat{
        \includegraphics[width=0.19\textwidth,valign=c]{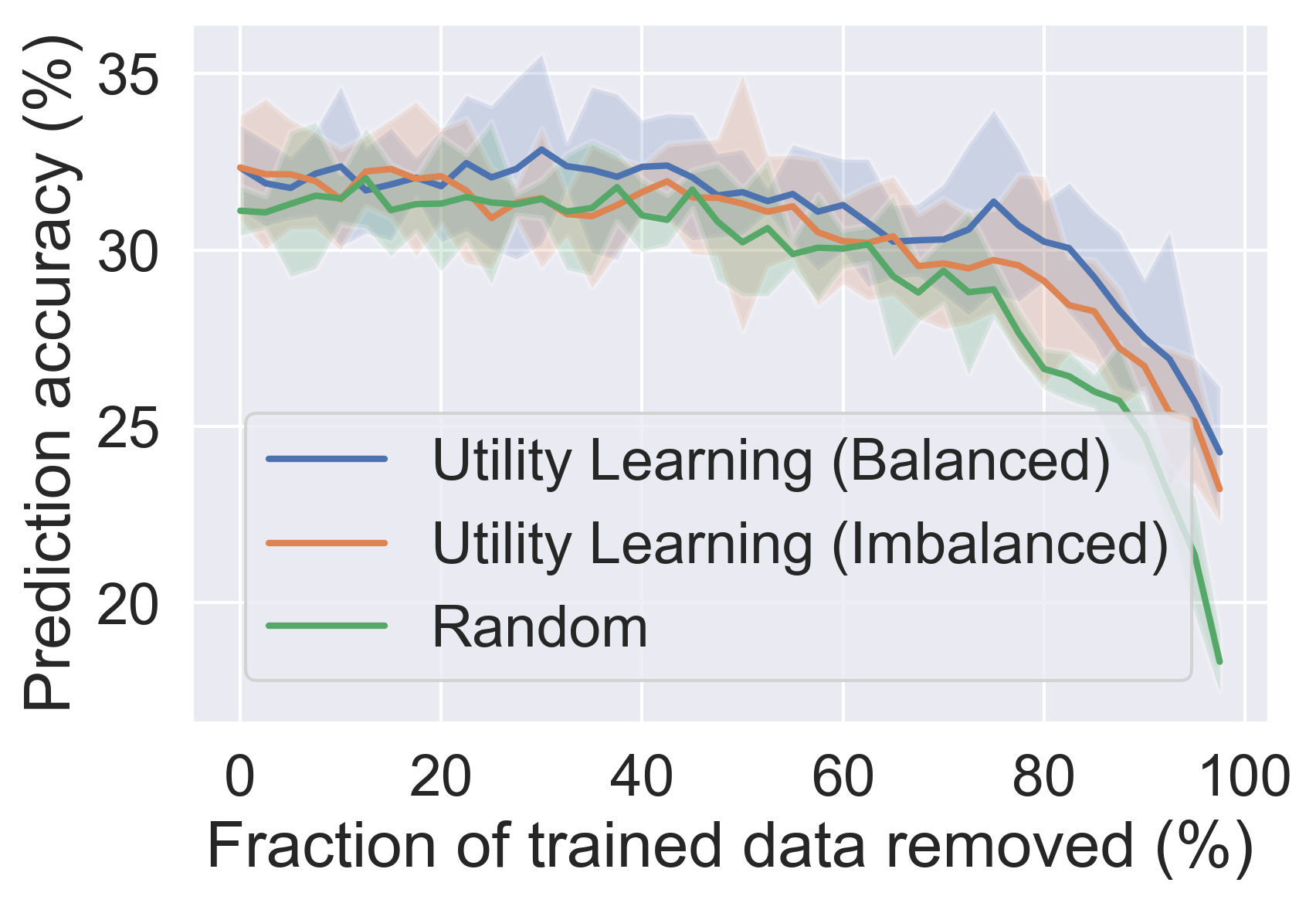}} &
    \subfloat{
        \includegraphics[width=0.19\textwidth,valign=c]{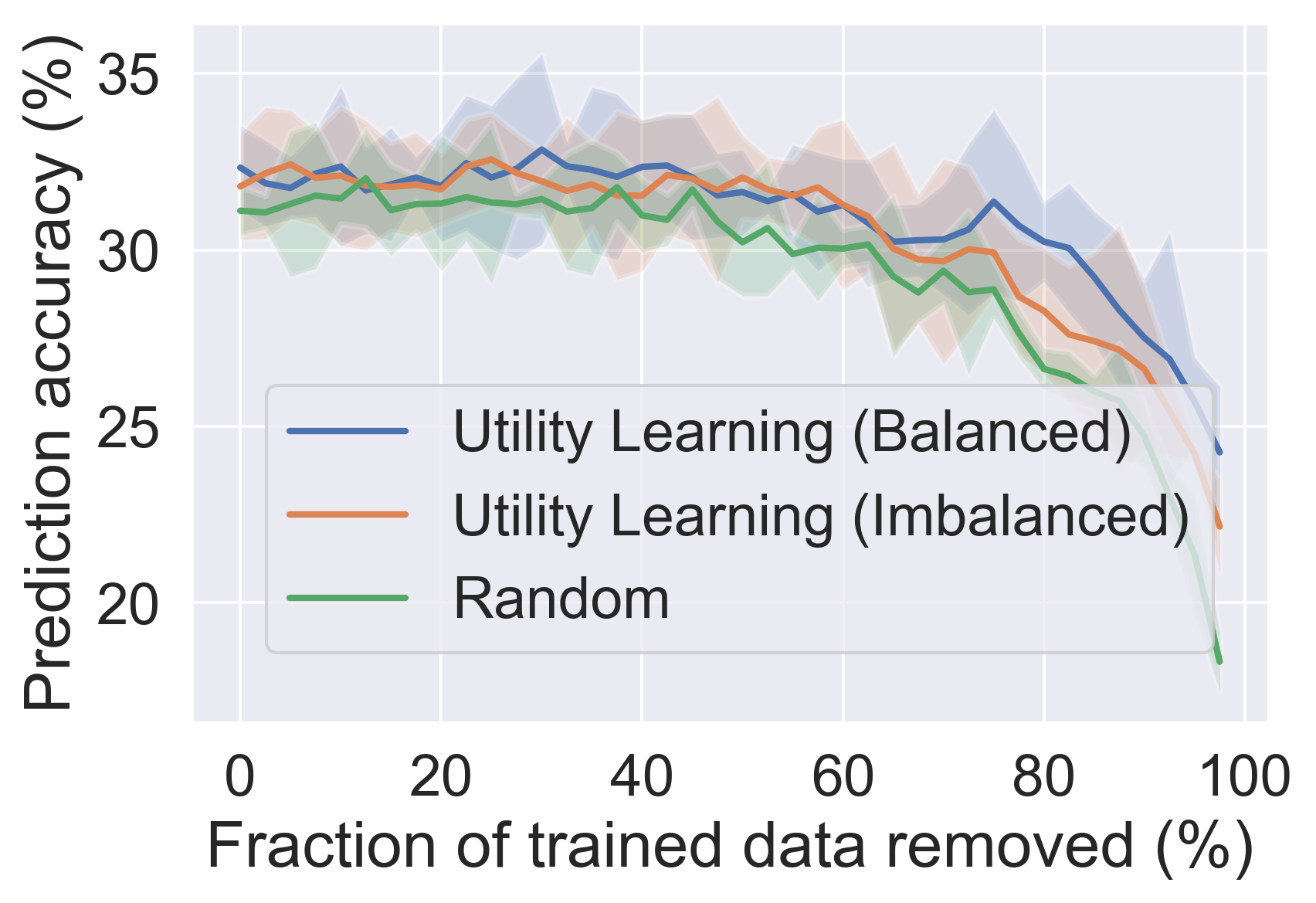}} \\
\addlinespace
    \rotatebox[origin=c]{90}{\parbox{2.5cm}{Removing high\\ value data}}&
    \subfloat{
        \includegraphics[width=0.19\textwidth,valign=c]{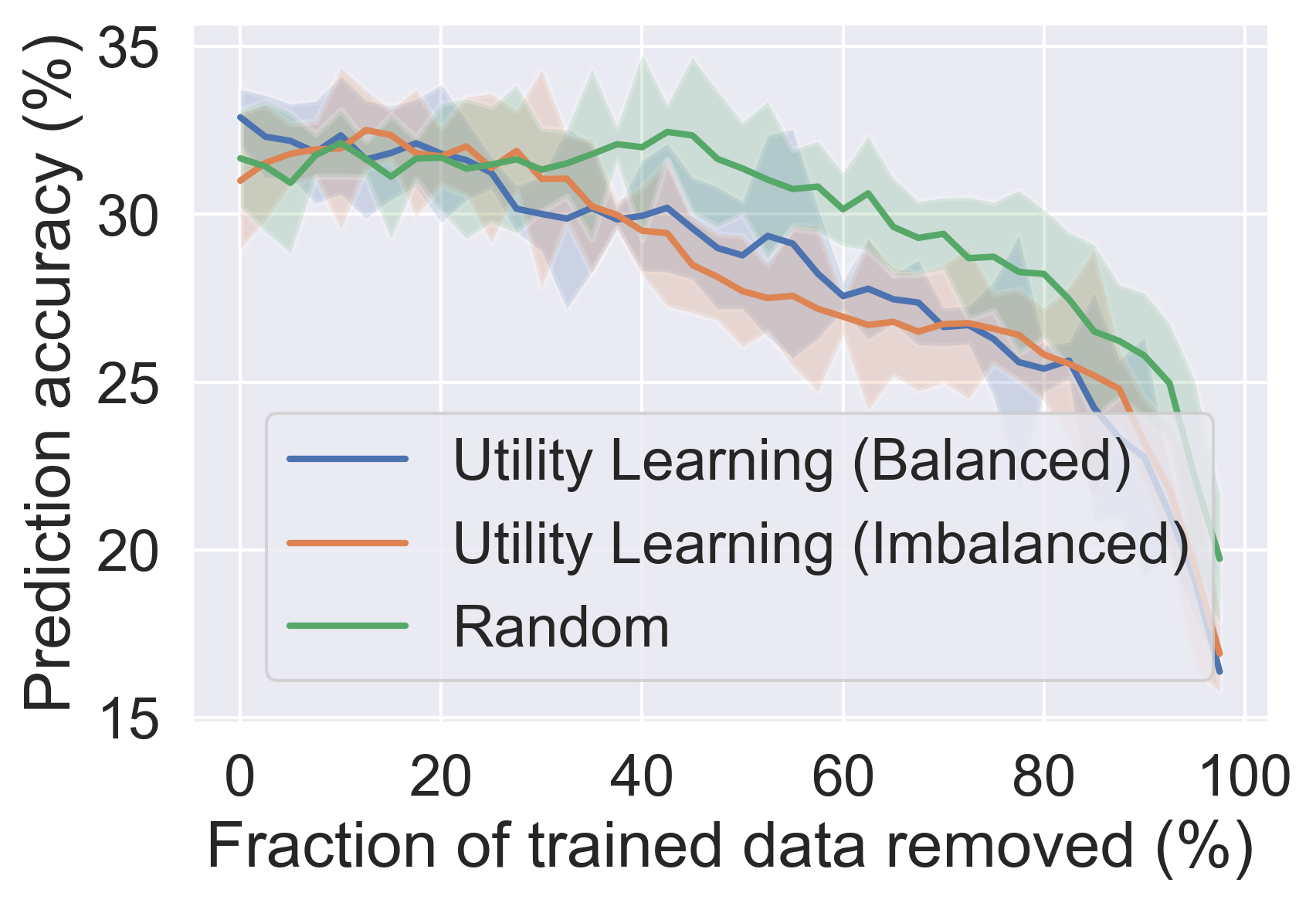}} &
    \subfloat{
        \includegraphics[width=0.19\textwidth,valign=c]{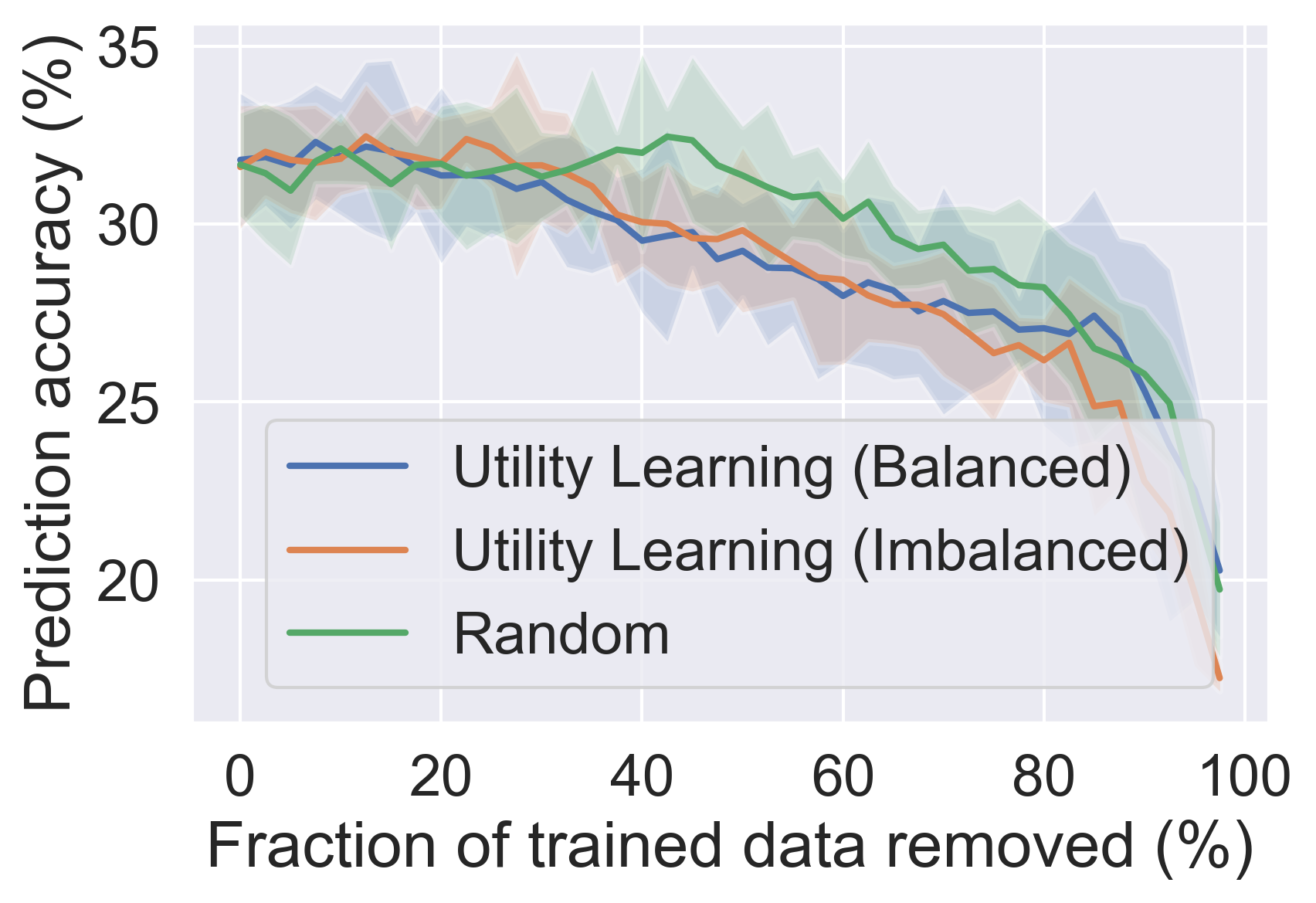}}
\end{tabular}
\caption{Models' validation accuracy vs. the proportion of removed instances for different settings. The accuracy line for the randomly removing strategy is generated with the balanced dataset.}
\label{fig:imbalance}
\end{figure}

\paragraph{\textbf{SV for Imbalanced Dataset}}
We artificially simulate data owners who own class-imbalance datasets for \textsl{CIFAR-10} and experiment with both the labeled pre-sharing method and the unlabeled pre-sharing method. We distribute data to data owners according to the Dirichlet distribution~\cite{Dirichlet}.
Specifically, for each class, we randomly generate a set of real values $\alpha \in [20, 100]$ with probability $0.2$ and generate a set of real values $\alpha \in [80, 100]$ with probability $0.8$. We then draw a distribution sample $p^i=(p_1^i,..,p_N^i)$ from Dirichlet distribution $\text{Dirichlet}(\alpha_1^i,..., \alpha_N^i)$ where $i$ represents the $i$-th class for \textsl{CIFAR-10} and $N$ represents the number of data owners.
We distribute $p_j^i$ to data owner $\Cli_j$ for data of class $i$. Finally, the proportion of class $j$ in $D_j$ is ${p_j^i}\slash{\sum_{k=1}^{10} p_j^k}$. In our experiments, we simulate 40 data owners, and each data owner holds a dataset with different sizes $|D_i|\in [300, 500]$. We also simulate 40 data owners who hold balanced datasets with the same size for experiments as comparison. 
The results for labeled pre-sharing method are shown in Figure~\ref{fig:imbalance}(a) and the results for unlabeled pre-sharing method are shown in Figure~\ref{fig:imbalance}(b). The accuracy line for imbalanced datasets is similar to that for balanced datasets, demonstrating that our methods are robust against imbalanced datasets.

\begin{table*}[!thb]
\centering
\resizebox{\textwidth}{!}{
\begin{NiceTabular}{c||c|c|c|c|c||c|c|c|c|c |c||c|c|c|c|c||c|c|c|c|c}  
\toprule 
 &\multicolumn{5}{c}{\textbf{Removing low value data}} &\multicolumn{5}{c}{\textbf{Removing high value data}}& &\multicolumn{5}{c}{\textbf{Removing low value data}} &\multicolumn{5}{c}{\textbf{Removing high value data}}\\\cline{2-11} \cline{13-22}
\textbf{setting} & \textbf{dataset size}& \multicolumn{4}{c}{\textbf{Number of data owners}} & \textbf{dataset size}& \multicolumn{4}{c}{\textbf{Number of data owners}}&\textbf{setting} & \textbf{dataset size}& \multicolumn{4}{c}{\textbf{Number of data owners}} & \textbf{dataset size}& \multicolumn{4}{c}{\textbf{Number of data owners}}\\
\midrule
\midrule
\multirow{5}{*}{a9a} &  & 10 & 20 & 30 & 40 & &10 & 20 & 30 & 40  &\multirow{5}{*}{MNIST} &  & 30 & 40 & 50 & 60 & &30 & 40 & 50 & 60 \\ \cline{2-11} \cline{13-22}
& 100 & 0.235 & 0.151 & 0.254 & 0.357 & 100 & 0.023 & 0.038 & 0.060 & 0.022 & & 100 & 0.043 & 0.016 & 0.033 & 0.024 & 100 & 0.089 & 0.120 & 0.181 & 0.283 \\  
& 200 & 0.276 & 0.161 & 0.358 & 0.262 & 200 & 0.017 & 0.038 & 0.043 & 0.008 & &200 & 0.052 & 0.014 & 0.010 & 0.027 & 200 & 0.190 & 0.158 & 0.255 & 0.156 \\ 
& 300 & 0.368 & 0.129 & 0.248 & 0.178 & 300 & 0.020 & 0.086 & 0.032 & 0.050 & &300 & 0.017 & 0.029 & 0.014 & 0.020 & 300 & 0.260 & 0.056 & 0.240 & 0.114 \\ 
& 400 & 0.274 & 0.378 & 0.255 & 0.160 & 400 & 0.048 & 0.051 & 0.028 & 0.034 & &400 & 0.017 & 0.004 & 0.033 & 0.016 & 400 & 0.109 & 0.260 & 0.079 & 0.128 \\ 
\midrule
\midrule
\multirow{5}{*}{CIFAR-10} &  & 10 & 20 & 30 & 40 & &10 & 20 & 30 & 40& \multirow{5}{*}{\makecell{MNIST$\Rightarrow$\\ USPS}} &  & 30 & 40 & 50 & 60 & &30 & 40 & 50 & 60 \\ \cline{2-11} \cline{13-22}
& 400 & 0.024 & 0.016 & 0.007 & 0.014 & 100 & 0.019 & 0.018 & 0.004 & 0.005 & & 100 & 0.063 & 0.059 & 0.043 & 0.015 & 100 & 0.024 & 0.009 & 0.013 & 0.044  \\ 
& 500 & 0.020 & 0.007 & 0.038 & 0.042 & 200 & 0.014 & 0.020 & 0.052 & 0.013 & & 200 & 0.017 & 0.008 & 0.028 & 0.025 & 200 & 0.019 & 0.028 & 0.014 & 0.074 \\ 
& 600 & 0.015 & 0.034 & 0.023 & 0.047 & 300 & 0.022 & 0.021 & 0.025 & 0.016 & & 300 & 0.014 & 0.024 & 0.046 & 0.008 & 300 & 0.041 & 0.019 & 0.059 & 0.006\\ 
& 700 & 0.004 & 0.046 & 0.027 & 0.013 & 400 & 0.039 & 0.032 & 0.029 & 0.026 & & 400 & 0.044 & 0.002 & 0.009 & 0.009 & 400 & 0.023 & 0.002 & 0.001 & 0.020\\ 

\bottomrule 
\end{NiceTabular} }
\caption{Experiments for the setting of different data owners and each data owner hold different number of data points.}
\label{tab:scalability}
\end{table*}

\paragraph{\textbf{SV for Adversarial Dataset}}
Given some small intentional feature perturbations, an adversarial example can cause a machine learning model to make a false prediction. 
However, training with adversarial examples could improve the robustness of an ML model. With more adversarial examples are added to the validation dataset, the adversarial dataset is expected to become more valuable. We artificially constructed a data owner who owns adversarial examples (we call such a data owner the \emph{adversarial data owner}), while other data owners own benign examples.
The public dataset contains only benign examples. 

All experiments are conducted under the setting of $8$ data owners each hold a dataset with \numprint{1000} instances. We vary the adversarial-benign mixing ratios to synthesize different validation datasets and observe the utility change of the adversarial data owner. We adapt the popular adversarial attack algorithm, namely Projected Gradient
Descent (PGD)~\cite{DBLP:conf/iclr/MadryMSTV18} to generate adversarial samples.
In more detail, we use the PGD-20 attack in our experiments. We set the $\epsilon=0.3$ for \textsl{MNIST} and \textsl{USPS} dataset and $\epsilon=0.031$ for \textsl{CIFAR-10} dataset~\cite{DBLP:conf/iclr/WongRK20}.

\begin{figure}[htbp]
\centering
\setlength\tabcolsep{1.0pt}
\begin{tabular}{ccc}
     & Labeled pre-sharing method& Unlabeled pre-sharing method\\[-0.5ex]
    \rotatebox[origin=c]{90}{\parbox{2.5cm}{\textbf{Shapley Value}}}&
            
    \subfloat{
        \includegraphics[width=0.22\textwidth,valign=c]{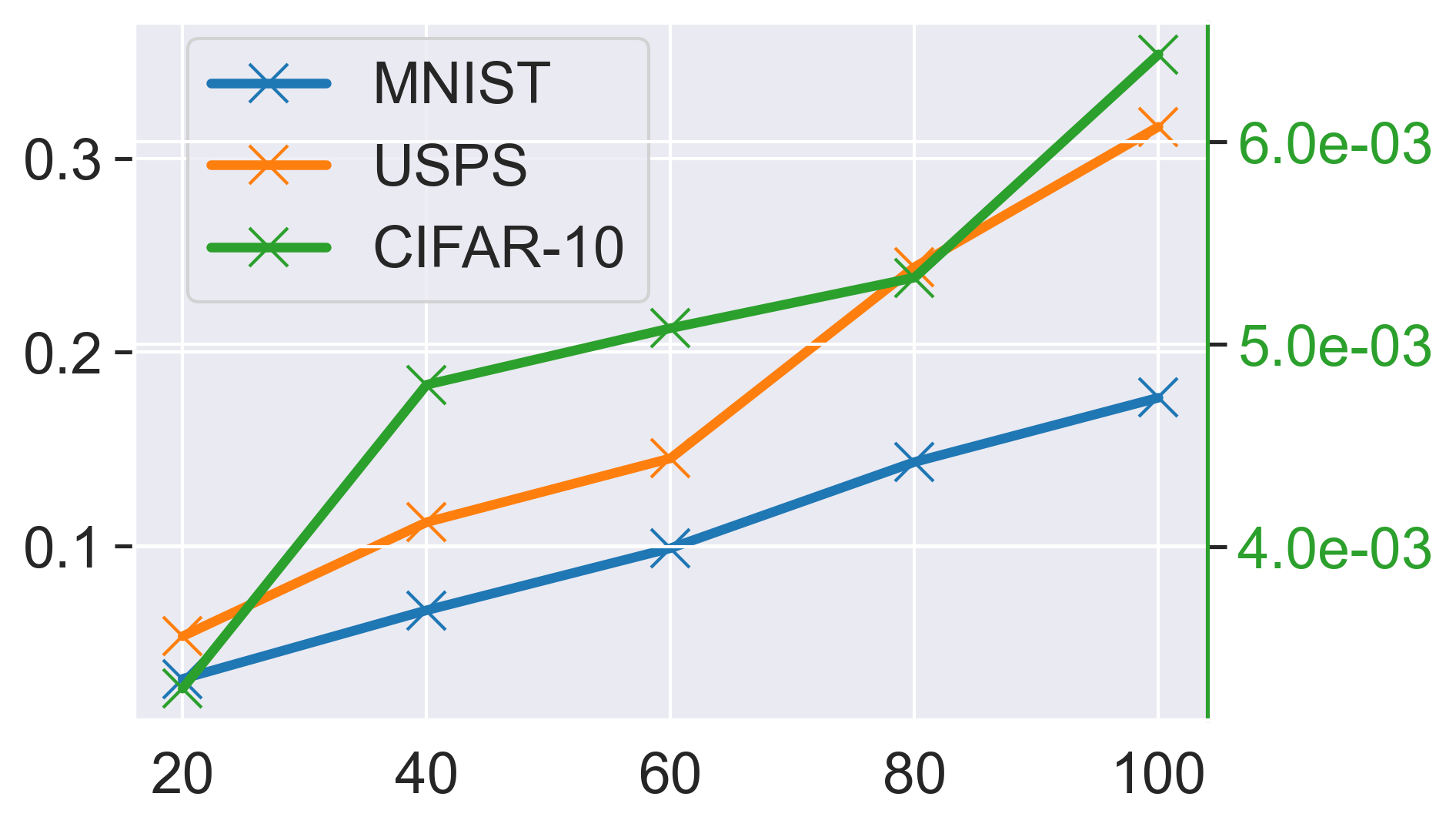}} &
    \subfloat{
        \includegraphics[width=0.22\textwidth,valign=c]{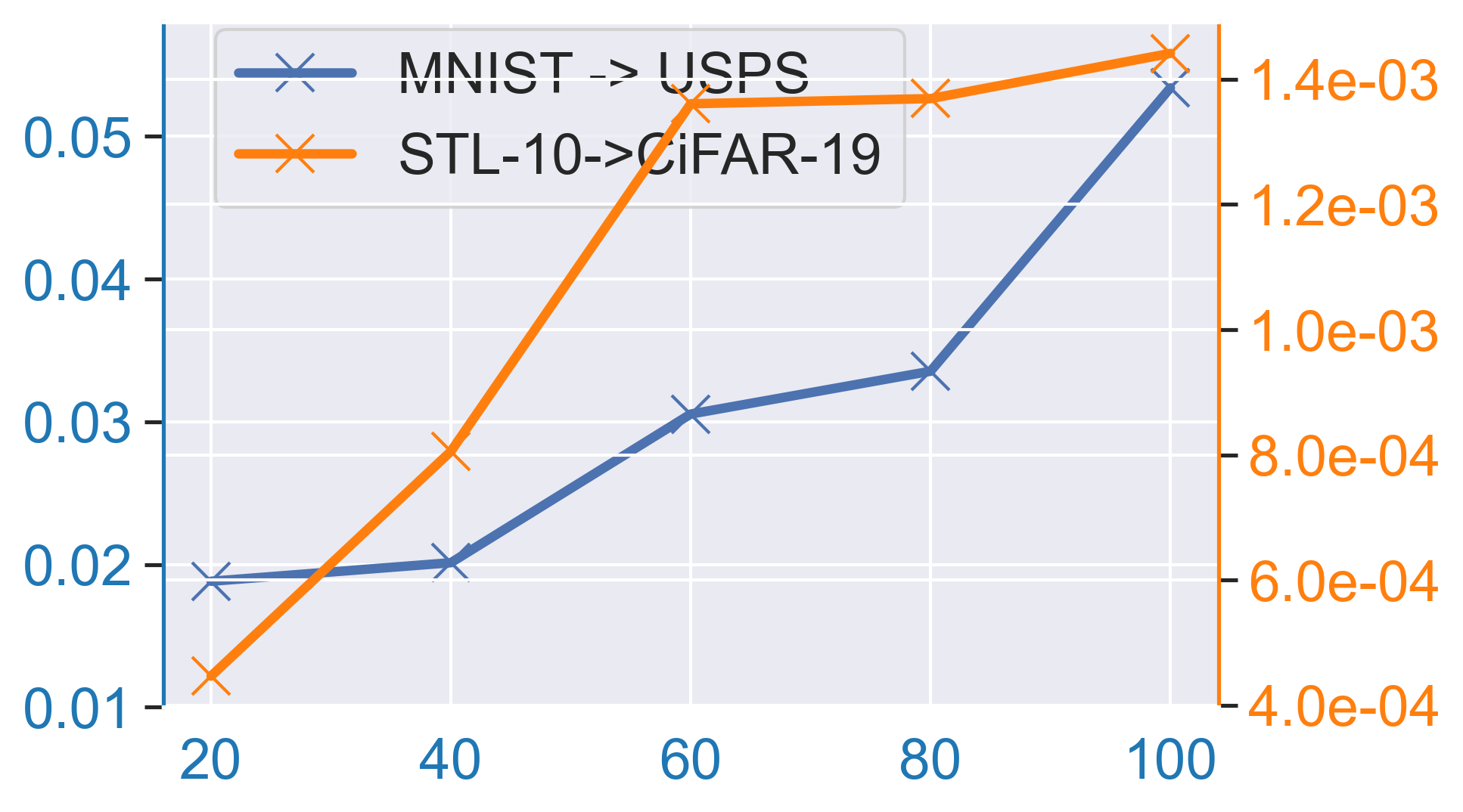}} \\
    & \multicolumn{2}{c}{\textbf{Proportion of adversarial samples in validation set (\%)}}

\end{tabular}
\caption{Shapley value for adversarial data owner VS. The proportion of adversarial samples in the validation set.}
\label{fig:adversarial}
\end{figure}

\begin{figure*}[htbp]
\centering
\setlength\tabcolsep{1.0pt}
\begin{tabular}{ccccc}
     & Labeled Pre-sharing & Unlabeled Pre-sharing & Labeled Pre-sharing &  Unabeled Pre-sharing\\
    \rotatebox[origin=c]{90}{\parbox{2.5cm}{\textbf{Shapley Value}}}&
            
    \subfloat{
        \includegraphics[width=0.23\textwidth,valign=c]{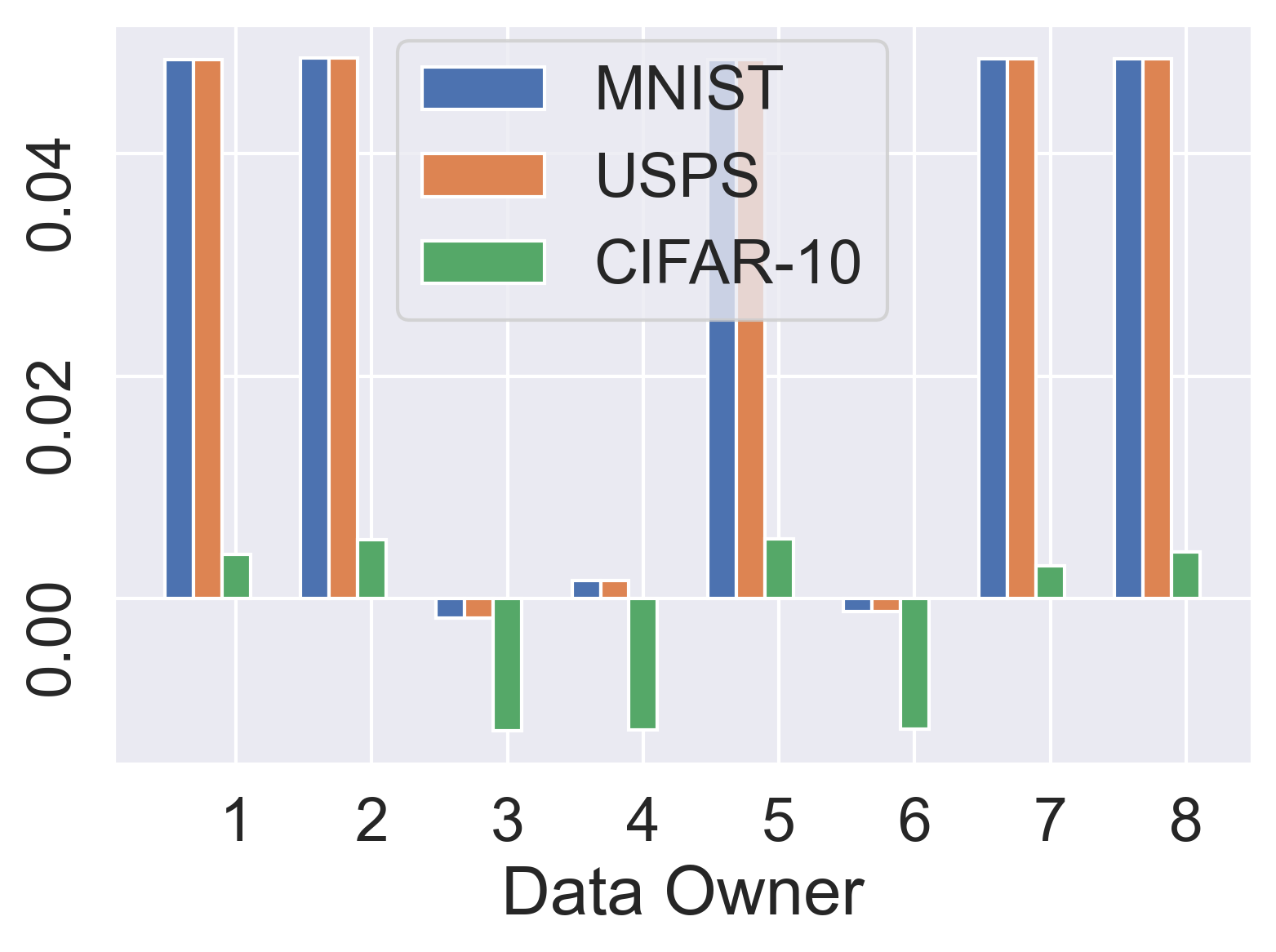}} &
    \subfloat{
        \includegraphics[width=0.23\textwidth,valign=c]{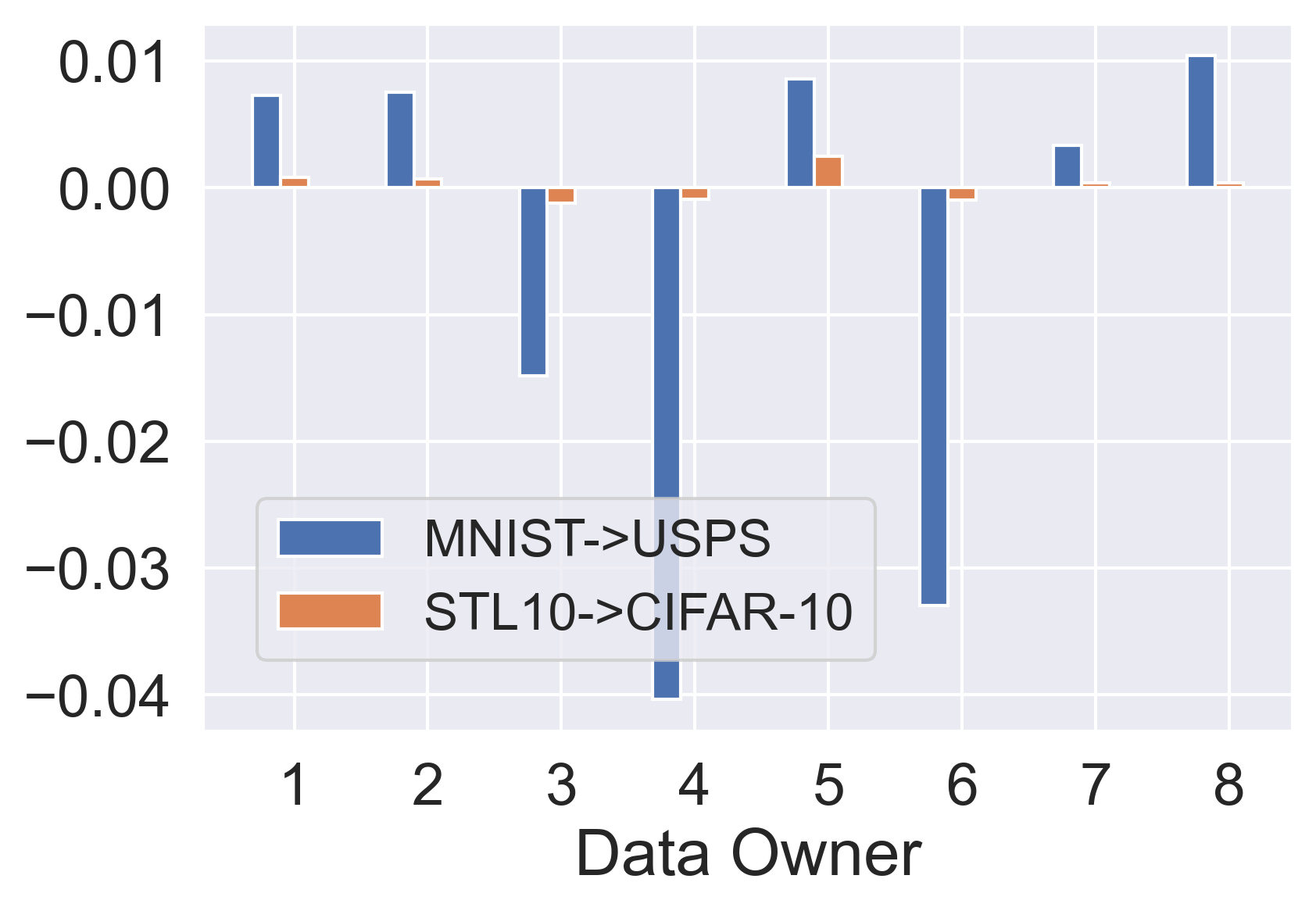}} &
    \subfloat{
        \includegraphics[width=0.23\textwidth,valign=c]{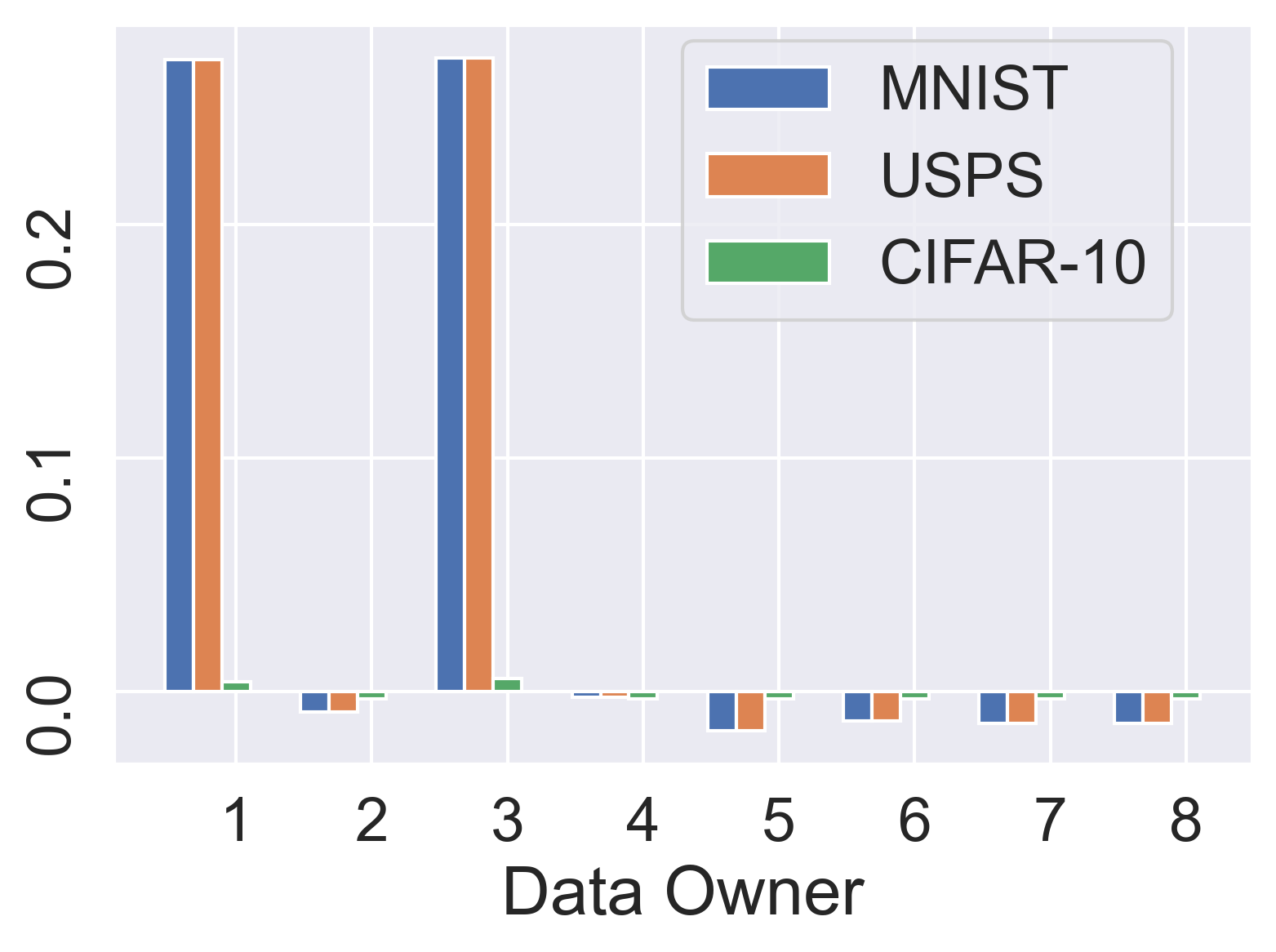}} &
    \subfloat{
        \includegraphics[width=0.23\textwidth,valign=c]{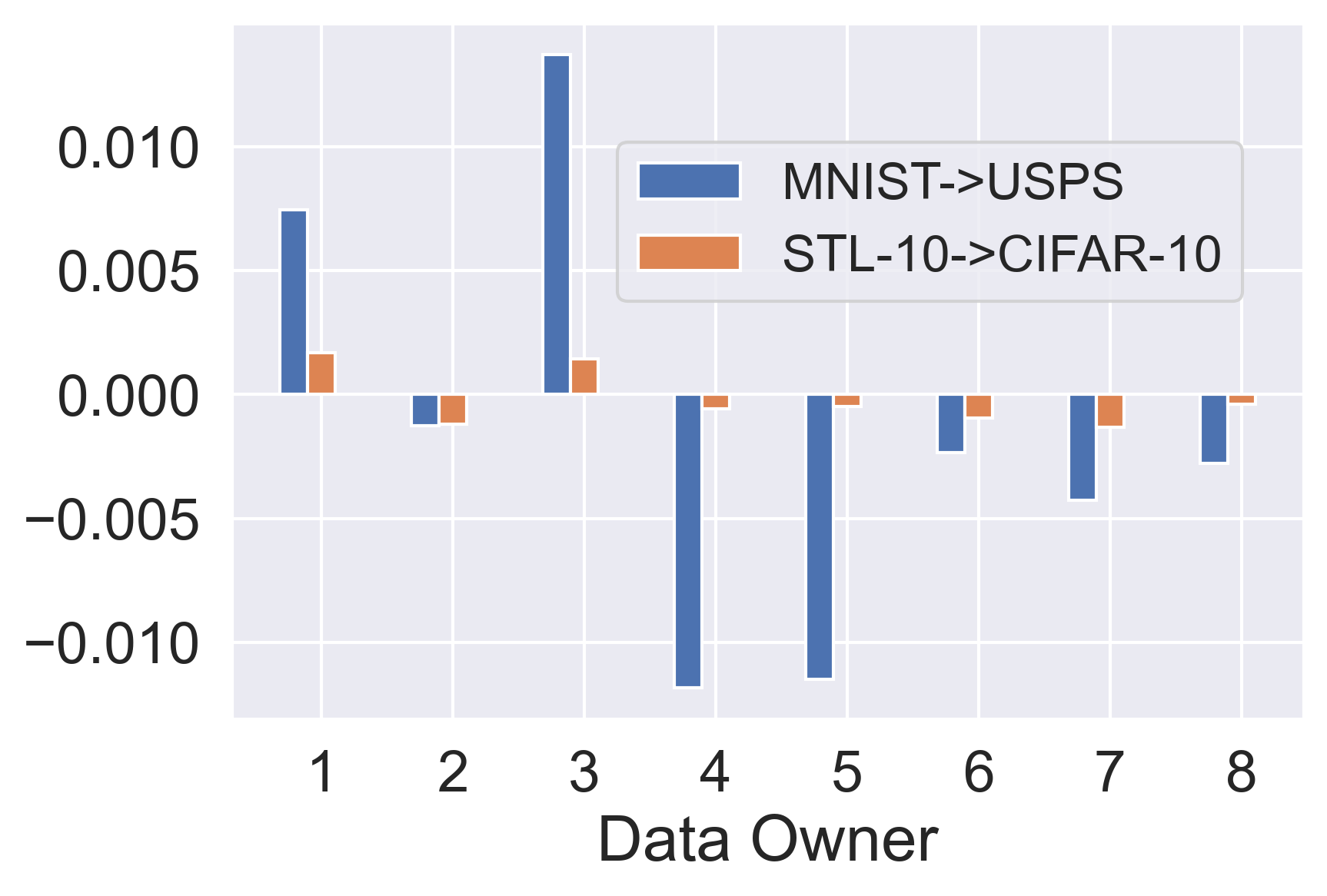}} \\
    & \multicolumn{2}{c}{\textbf{(a) Specify data owners $\Cli_3$, $\Cli_4$, $\Cli_6$ as dishonest}} &
    \multicolumn{2}{c}{\textbf{(b) Specify data owners $\Cli_2$, $\Cli_4$, $\Cli_5$, $\Cli_6$, $\Cli_7$, $\Cli_8$ dishonest}}

\end{tabular}
\caption{Shapley value for data owners. Dishonest data owners who intentionally select good data to pre-share.}
\label{fig:dishonest}
\end{figure*}

The results are presented in Figure~\ref{fig:adversarial}.
It shows that the SV of the adversarial data owner increases as the validation data become more adversarial, consistent with our expectations. 
Meanwhile, the SV for \textsl{CIFAR-10} is smaller than \textsl{MNIST} and \textsl{USPS}, because the learning algorithm $\mathcal{A}$ for \textsl{CIFAR-10} is unstable during the training dataset construction phase. In other words, the validation accuracy of datasets with different utilities may be close, resulting in the data utility model cannot distinguish the utility of datasets well. However, such problem can be resolved by running the learning algorithms multiple times or adopting a more robust learning algorithm when constructing the training set $\LL{tr}$.

\paragraph{\textbf{SV for Intentionally Selected Dataset}}
Recall that each data owner pre-shares a proportion of data with the buyer, who exploits these data to train the data utility model.  
However, what if data owners who own mostly low-SV data intentionally select high-SV data for pre-sharing to bias the distribution of the training set? To verify the robustness of our method against such activity, we conduct experiments to artificially simulate such data owners and verify the robustness of our methods. For simplicity, we call data owners who intentionally select good data to pre-share as \emph{dishonest data owner} and those who randomly select data to pre-share as \emph{honest data owner}. 
We simulate $8$ data owners each holding a dataset with \numprint{1000} instances. Dishonest data owners are simulated by adding Gaussian noise with $\sigma=10$ to its data, but still pre-share clean data without noise. We also vary the number of dishonest data owners to verify the performance of our methods under majority dishonest, i.e., most of the data owners are dishonest. 

Figure~\ref{fig:dishonest} (a) presents the SV for minority dishonest data owners, and Figure~\ref{fig:dishonest} (b) presents the SV for majority dishonest data owners. It shows that all dishonest data owners with low-SV data have negative SV, and honest data owners with high-SV data have positive SV. We conjecture it is probably because the data utility model can still learn the useful feature from the natural variance of data in the distribution. Furthermore, the results also show that the unlabeled pre-sharing method is more sensitive for dishonest data owners than the labeled pre-sharing method, probably because the data in the source domain enrich the variety of training data for the data utility model.

\paragraph{\textbf{SV for Noisy Labeled Dataset}}
We then conduct experiments to explore the influence of noisy labels of the labeled pre-sharing method with the \textsl{MNIST} dataset.
For a label that is originally assigned as class $i$, we add noise using the following strategy: 
\begin{itemize}
    \item with probability $p$, it keeps unchanged;
    \item with probability $1-p$, it changed to the $j$th class that is picked uniformly at random. 
\end{itemize}


We simulate $60$ data owners and each data owner has a class-balanced data set of $300$ instances.
Figure~\ref{fig:noisy_label} shows the changes in the effectiveness score~\ref{equation:effectiveness_score} when $p$ ranges from $0.6$ to $1.0$.
It shows that mislabeled data do degrade the effectiveness of the utility model. However, our method can still evaluate the utility of data even if $40\%$ data points are mislabeled. 
Meanwhile, the effectiveness score of removing high-SV data bigger than that of removing low-SV data, probably because high-SV data contributed more to train a better model, and removing more high-value has a greater impact on accuracy.

\subsection{Proportion of Pre-shared Data}
\label{ssec:shared_data}
The number of pre-shared data needs to be appropriately large to train a data utility model that can generalize well to unseen data points and subsets.
However, pre-sharing too much data to the buyer increases privacy leakage risk while having no compensation. 
To avoid such threats for data owners, training a practical data utility model while pre-sharing as little data as possible is needed.

\begin{figure*}[!ht]
\centering
\setlength\tabcolsep{1.0pt}

\begin{minipage}[b]{0.25\textwidth}
    \centering
    \includegraphics[width=1\textwidth,valign=c]{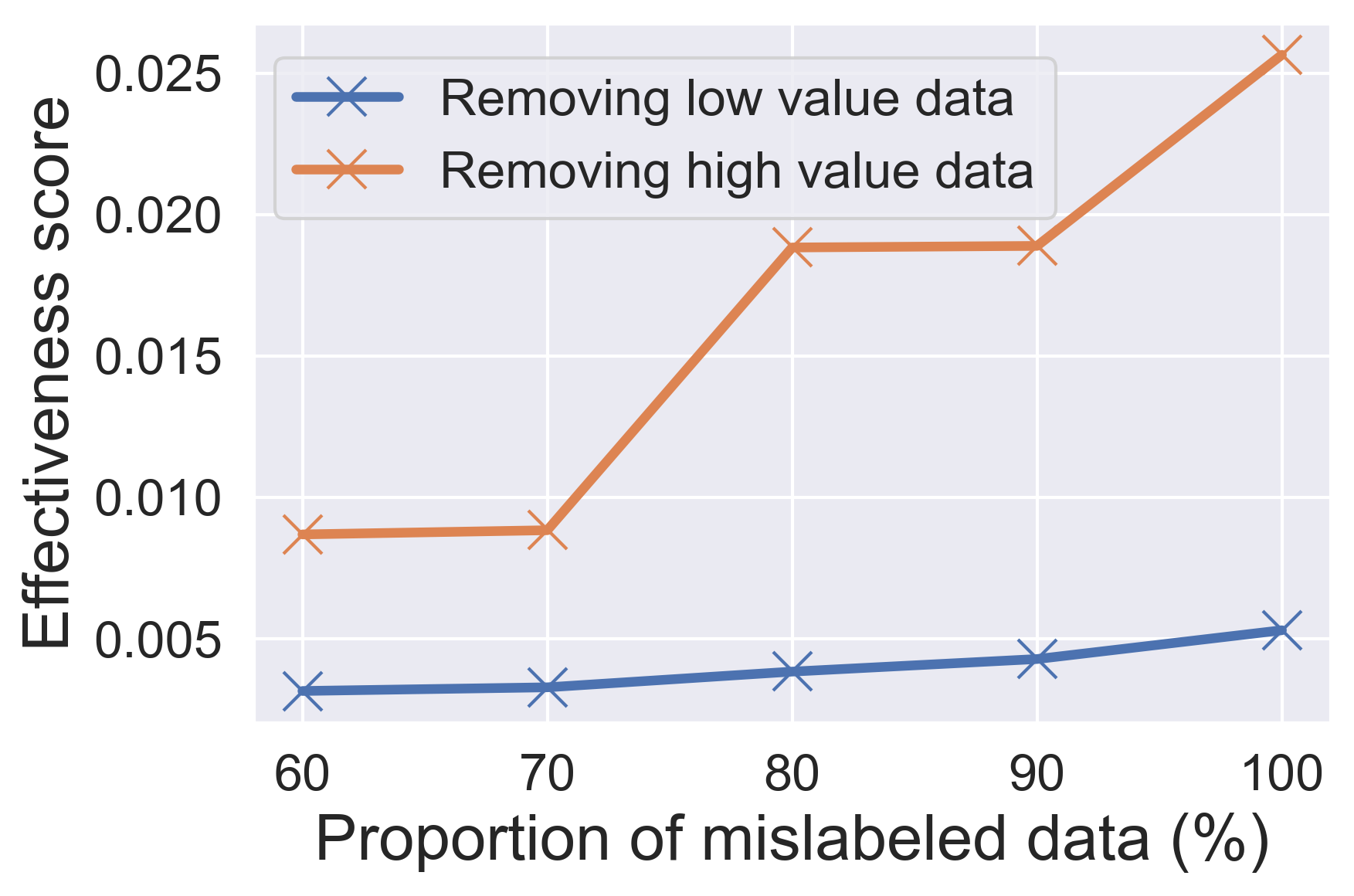}
    \caption{Effectiveness score VS. The proportion of correctly labeled data}
    \label{fig:noisy_label}
\end{minipage}
\begin{minipage}[b]{0.72\textwidth}
\begin{tabular}{ccc}
    Removing low value data & Removing high value data & \\[-0.5ex]
    \subfloat{
        \includegraphics[width=0.33\textwidth,valign=c]{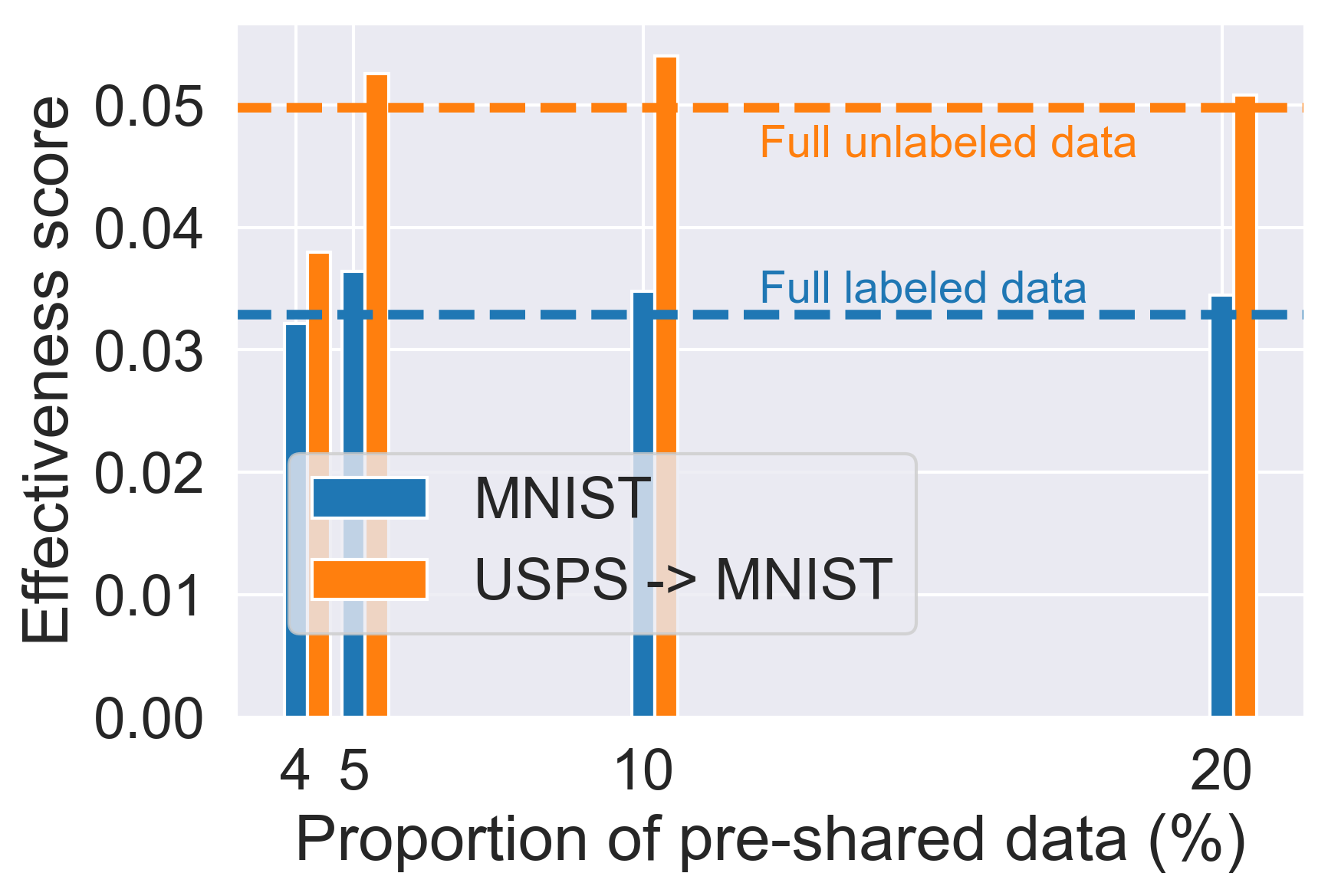}} &
    \subfloat{
        \includegraphics[width=0.33\textwidth,valign=c]{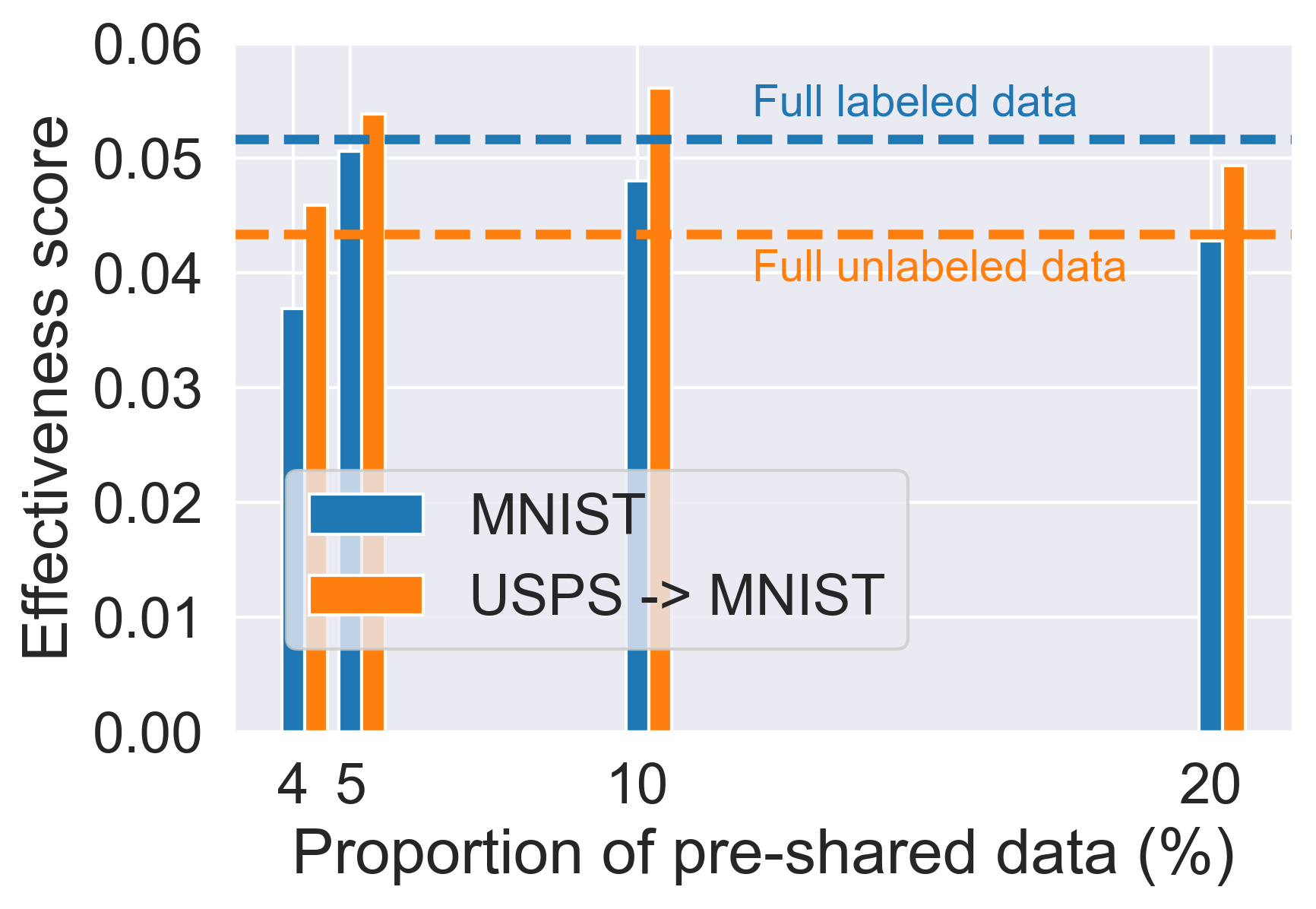}}& 
    \subfloat{
        \includegraphics[width=0.33\textwidth,valign=c]{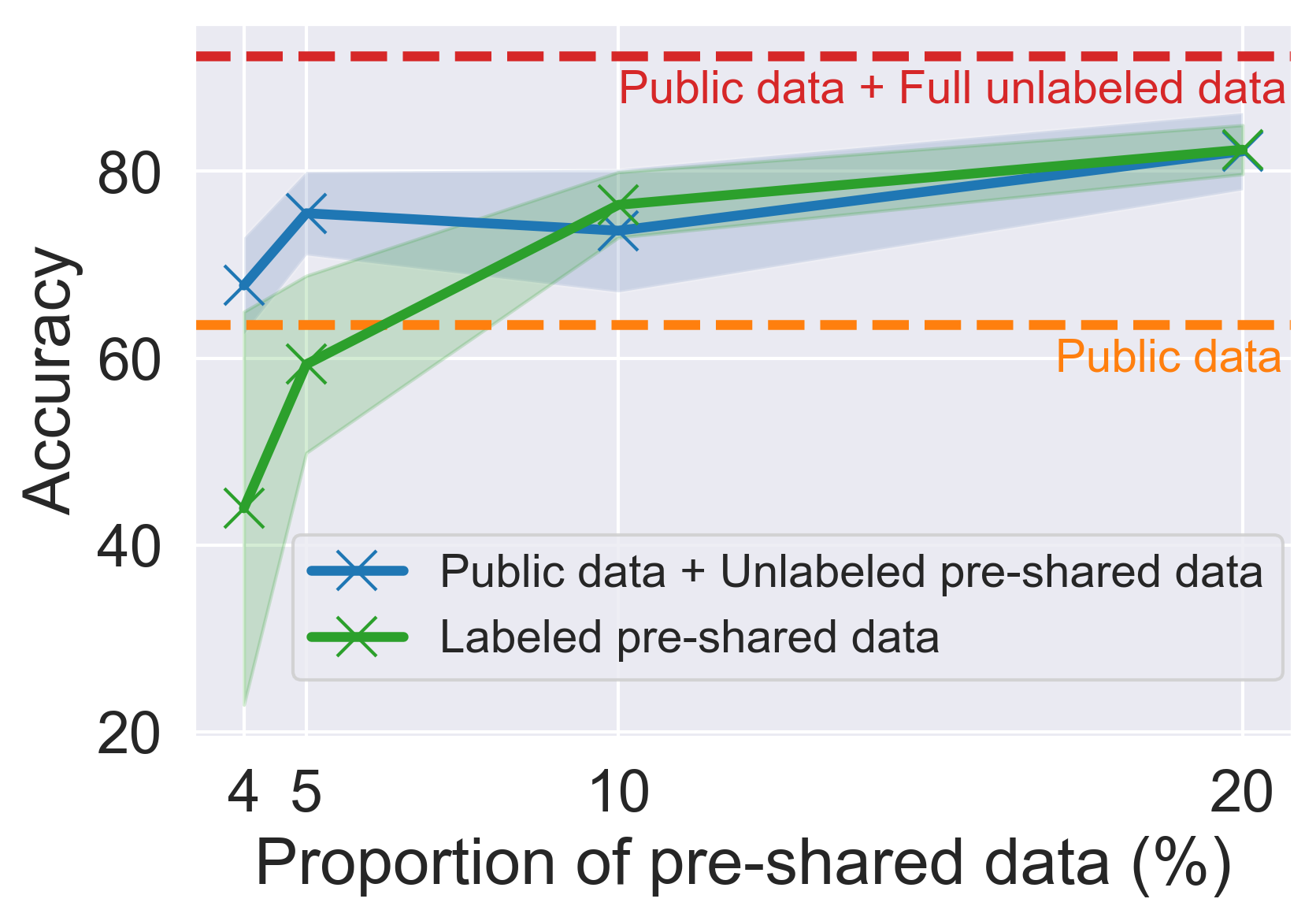}} \\
     \multicolumn{2}{c}{(a)} & \multicolumn{1}{c}{(b)}
\end{tabular}
\captionsetup{width=0.7\textwidth}
\captionof{figure}{(a) Effectiveness score VS. The proportion of pre-shared data, (b) Marginal contribution of pre-share data.}
\label{fig:share}
\end{minipage}
\end{figure*}

Figure~\ref{fig:share}(a) depicts the results of pre-sharing different proportion of data to the buyer. 
We experiment with the \textsl{MNIST} dataset for the labeled pre-sharing method and \textsl{USPS}$\Rightarrow$\textsl{MNIST} for the unlabeled pre-sharing method. We simulate $10$ data owners, each of them holds $100$ instances. The proportion of pre-shared data ranged from $4\%$ to $20\%$. We find that pre-sharing more data do perform better in training a practical utility function. However, only pre-sharing 5\% data is still enough, which means each data owner is only required to pre-share five samples with the buyer.
Meanwhile, we also conduct experiments to compare the utility of data that the buyer gets without extra costs (i.e., the public dataset and the received pre-shared data) with all those data that she intends to buy.
We train models with four different datasets:
\begin{enumerate}
    \item public dataset,
    \item labeled pre-shared dataset, which is consistent with our labeled setting, 
    \item public dataset plus unlabeled pre-shared dataset, which is consistent with our unlabeled setting,
    \item public dataset plus full unlabeled data.
\end{enumerate}
The results are depicted in Figure~\ref{fig:share} (b). In the labeled setting where the buyer only has the validation set, training with 5\% of pre-shared data can only train a model with about 60\% accuracy. In the unlabeled setting where the buyer has access to a public dataset, with an extra 5\% of pre-shared data, the buyer trains a model of 75\% accuracy and about 10\% of marginal contribution beyond only training with the public dataset. However, the marginal contribution of complete data can achieve above 35\% accuracy.
Combined with the results in Figure~\ref{fig:share}(a), it shows that we can train a utility function with data of low marginal contribution over what the buyer owns and the public dataset.


\subsection{Experiments for MPC circuit}
\label{sec:experiments_for_mpc}
We evaluate the performance of the MPC circuit for SV calculation, data encryption, and hashing.
Recall that we optimize our MPC circuit by replacing partial MPC circuit with 2PC circuit (cf. Section~\ref{ssec:optimization_mpc_circuit}). The data encryption and key hashing process is also implemented using a 2PC circuit. For simplicity, we name the process of running 2PC circuit as \textbf{2PC part} and running MPC circuit as \textbf{MPC part}.
We start by analyzing the 2PC part which takes most of the time when the number of data owners is relatively small.
\subsubsection{Performance of the 2PC Part}
\begin{figure}[htbp]
\centering
\setlength\tabcolsep{1.0pt}
\begin{tabular}{cc}
      MNIST & CIFAR-10\\[-0.5ex]
    \subfloat{
        \includegraphics[width=0.24\textwidth,valign=c]{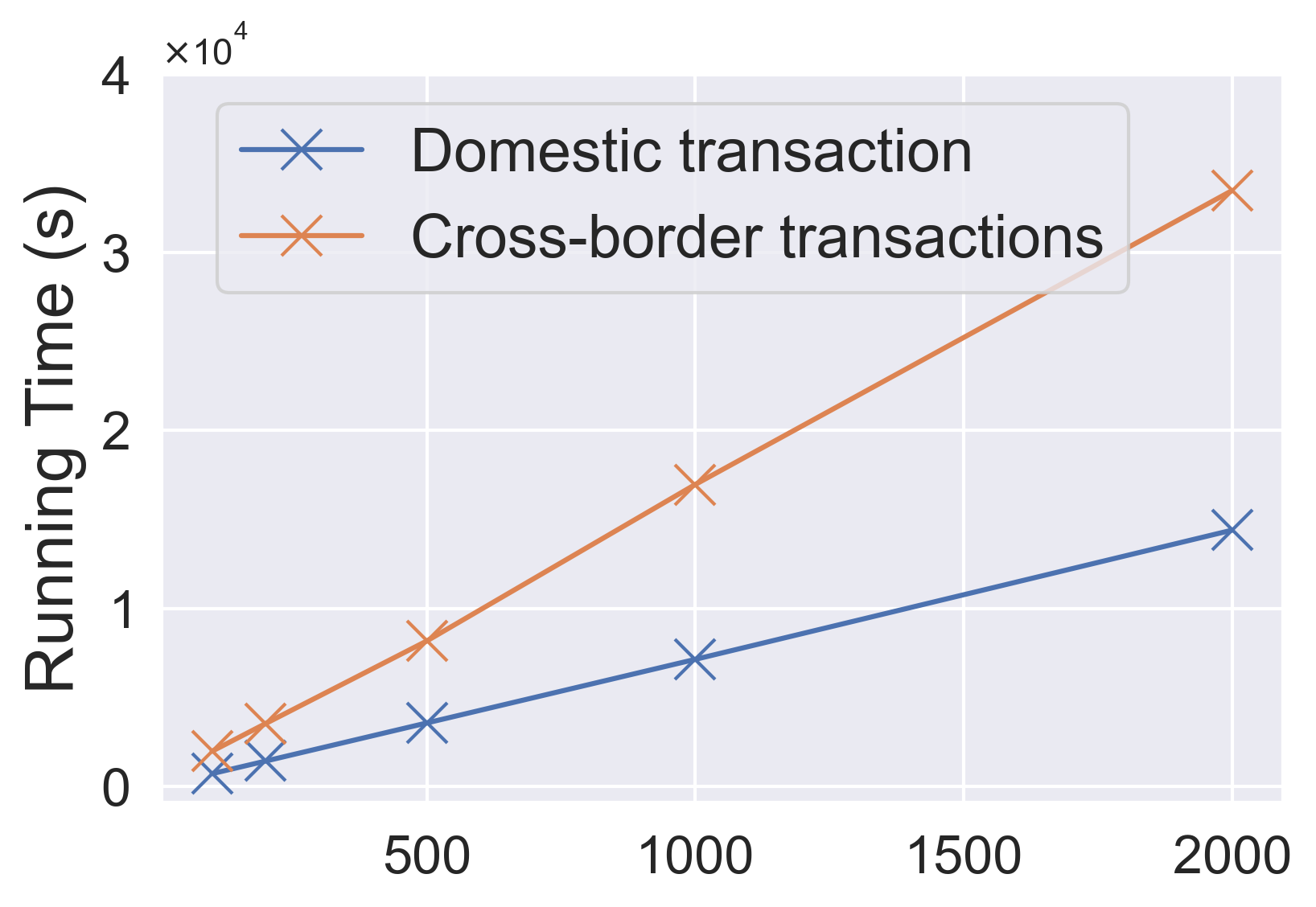}} &
    \subfloat{
        \includegraphics[width=0.24\textwidth,valign=c]{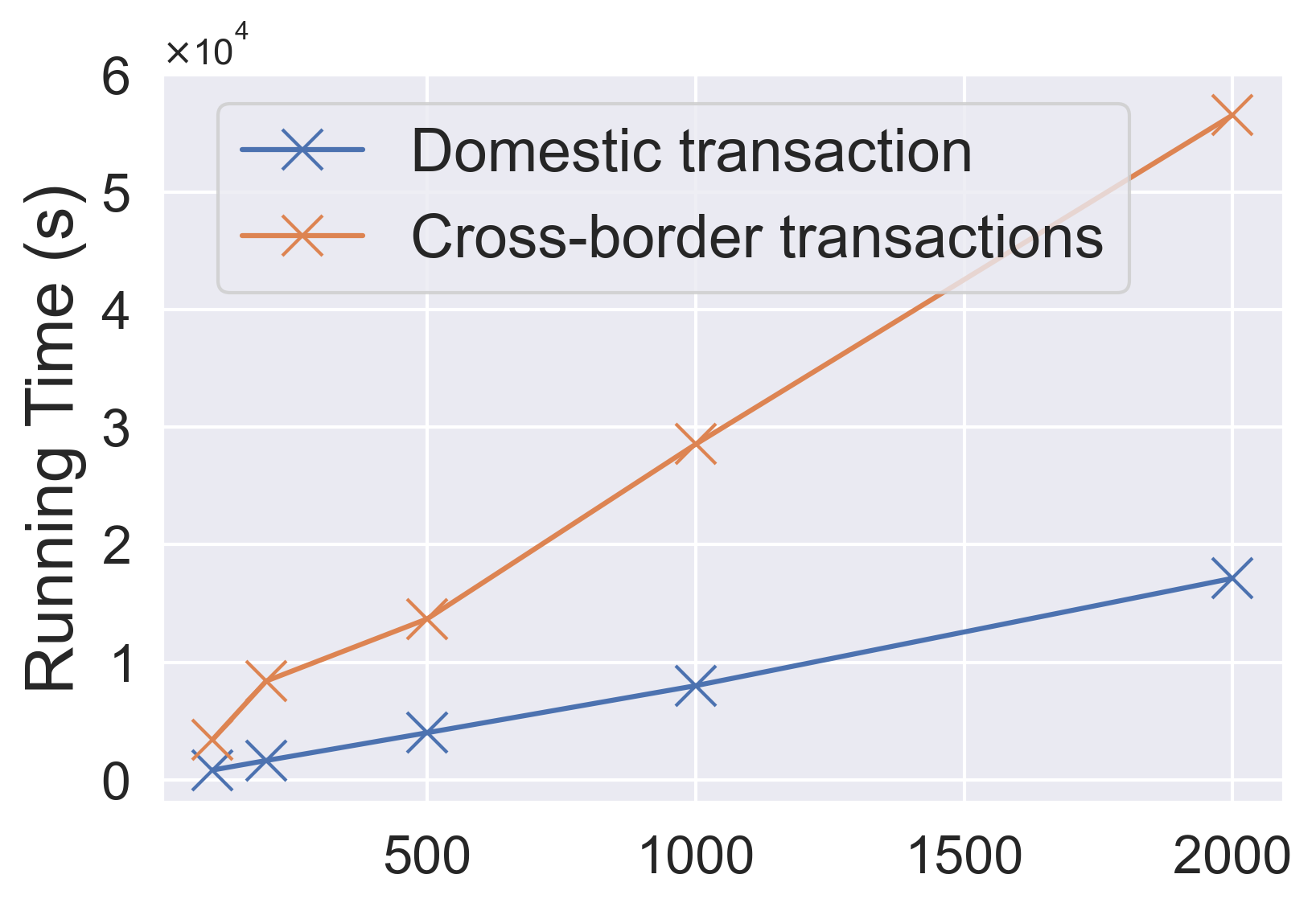}} \\
    \multicolumn{2}{c}{\textbf{Number of data points held by each data owner}}
\end{tabular}
\caption{Running time of 2-PC protocol VS the number of instances}
\label{fig:2pc_time}
\end{figure}

We use AES-256 as the encryption algorithm and SHA-256 as the hash algorithm, which is widely used on the blockchain for fair payment. 
We only evaluate the performance of the online phase of the SPD$\mathbb{Z}_{2^k}$ protocol in our experiments.
\paragraph{\textbf{Running time }}
Figure~\ref{fig:2pc_time} shows the running time of 2PC circuits with different number of data points for \textsl{MNIST} and \textsl{CIFAR-10}. As the prediction processes of utility models trained by labeled pre-sharing method and the unlabeled pre-sharing method are the same, we do not distinguish them here. The running time almost linearly grows with the number of instances. The 2PC part in a cross-border transaction, which is more time-consuming than a domestic transaction, costs no more than 10 hours for \numprint{2000} \textsl{MNIST} instances and 16 hours for \numprint{2000} \textsl{CIFAR-10} instances. In a domestic transaction, running the 2PC part for \numprint{2000} \textsl{MNIST} instances takes no more than 4 hours. For \numprint{2000} \textsl{CIFAR-10} it costs no more than 5 hours.

\begin{figure}[htbp]
\centering
    \subfloat[]{
        \includegraphics[width=0.24\textwidth,valign=c]{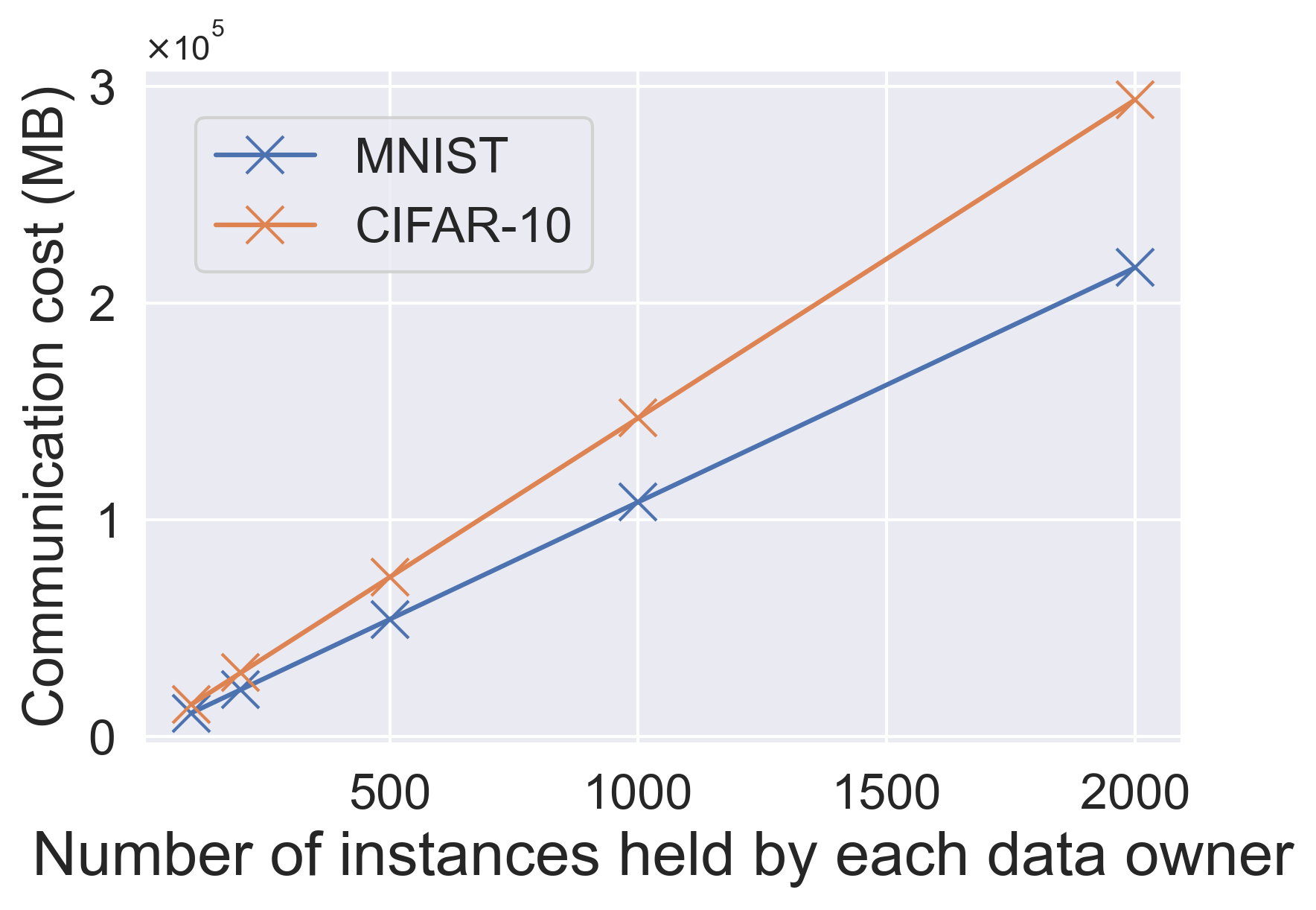}} 
    \subfloat[]{
        \includegraphics[width=0.24\textwidth,valign=c]{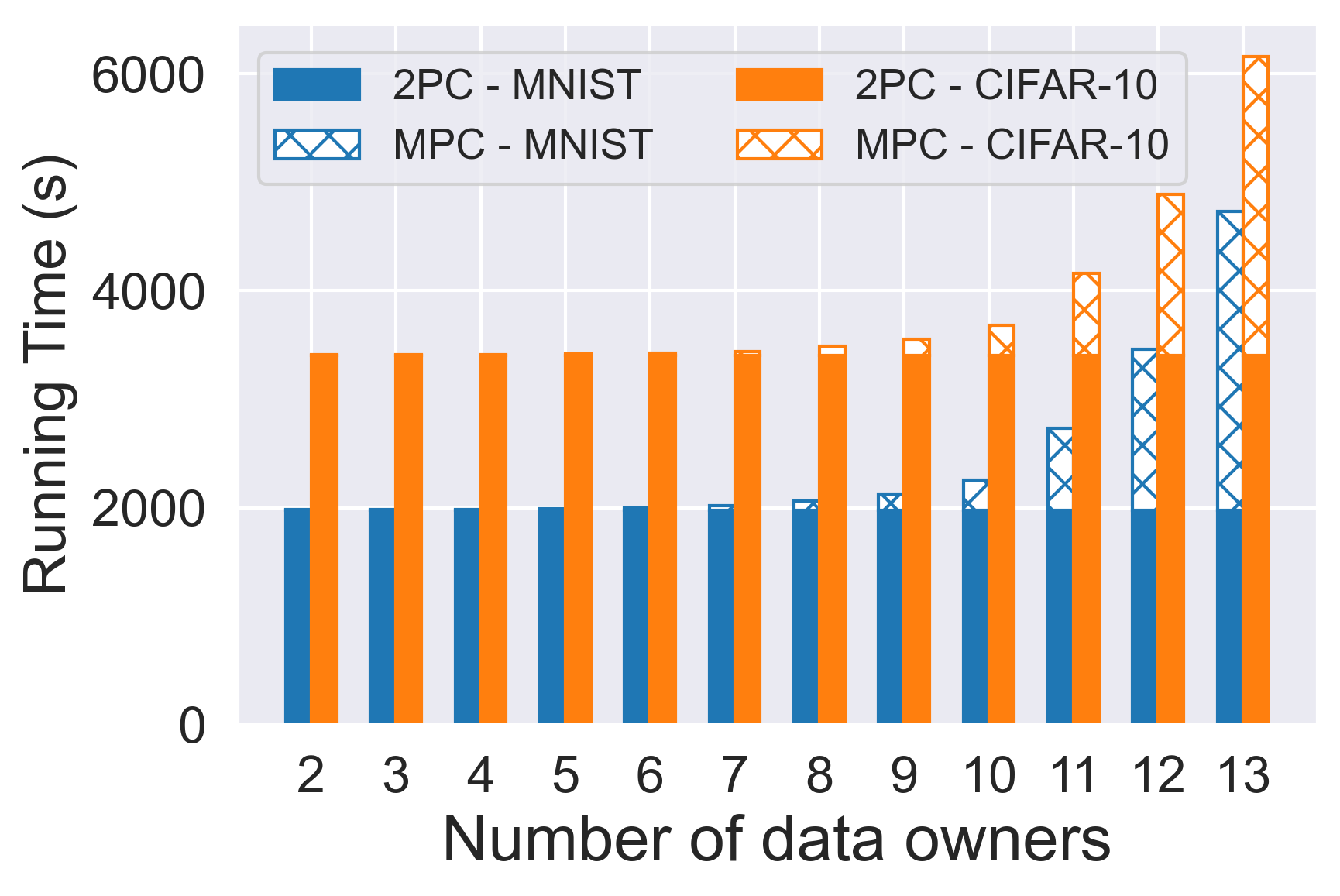}} 
\caption{(a) Communication cost of the 2PC circuits; (b) Running time of the 2PC and MPC parts.}
\label{fig:communication_cost}
\end{figure}

\paragraph{\textbf{Communication cost}}
We also evaluate the communication cost of the 2PC part. The results is shown in Figure~\ref{fig:communication_cost}(a). It show that communication cost grows linearly with the number of instances, which corresponds to the change of running time in Figure~\ref{fig:2pc_time}.

\subsubsection{Performance of the MPC Part}
The MPC part consists of the remaining prediction process of the data utility model (i.e. Average and $f_{DS}^{network}$), which mainly consists of a few fully-connected layers. Figure~\ref{fig:communication_cost} (b) depicts the results of the running time for 2PC and MPC for 100 instances.
We find that when the number of participants is small (e.g. less than 8), the running time of MPC is negligible. Moreover, the running time of the MPC part will be longer than running the 2PC part only when achieving 13 participants. It is worth noting that the running time of the MPC part would not change with the number of data growing since the Average operation map all inputs to a vector of the same shape. 
When the number of instances grows, the running time of the MPC part will take a smaller percentage.
This means that the running time of our circuits almost unaffected by the number of data owners.


\section{Related work}
\label{sec:related_work}
Data commoditization has recently become an emerging trend. The problem of data pricing, especially, gains a lot of attention. The pricing schemes currently deployed in the data marketplace are simplistic: a buyer either buy the whole or parts of the dataset for a fixed price. Existing data marketplace platforms like Datarade~\cite{Datarade} sells data based on the data type, size, query frequency, etc. Another kind of platforms like Google BigQuery~\cite{BigQuery} also charging for running (relational) queries over the dataset.

A bunch of previous works have studied how to price data over different kinds of data marketplaces. A recent line of works~\cite{DBLP:journals/pvldb/LinK14,DBLP:journals/jacm/KoutrisUBHS15} have formally studied pricing schemes for fine-gained queries over a dataset. With the query-based pricing schemes, given a dataset $D$ and a query $Q$, the seller assigns a price $p(D, Q)$ based on the information disclosed by the query answers. 
Two important properties that the pricing function must satisfy namely arbitrage-freeness and discount-freeness was proposed by Koutris el al~\cite{DBLP:journals/jacm/KoutrisUBHS15}. The arbitrage-freeness indicated that when query $Q_1$ discloses more information than $Q_2$, we want to ensure that $P(Q_1)>p(Q_2)$; otherwise, buyers have an arbitrage opportunity to take advantage of the marketplace by combining datasets for lower prices into a high-price dataset to escape the designated price for that dataset. The discount-freeness requires that the prices offers no additional discounts than the ones specified by the data seller. In fact, discount-freeness is the discrete version of arbitrage-freeness. 
Menwhile, with the increasing pervasiveness of machine learning-based analysis, model-based pricing has become a new interest in studying the cost of acquiring data for machine learning. Chen et al.~\cite{DBLP:conf/sigmod/ChenK019} proposed the first model-based pricing framework that directly prices instances of the ML model with different noise, in which an optimization problem was formulated to find the arbitrage-free price that maximizes the revenue of a seller. Liu et al.~\cite{DBLP:journals/pvldb/LiuLL0PS21} proposed an end-to-end model marketplace aiming to maximize the revenue for data owners while supplying the demands of model buyers, wherein a broker is introduced to collect data from data owners and sell ML models to buyers. 

The problem of fairness in the data marketplace also attracts researchers deeply, wherein Shapley value has become the widely used notion of fairness due to its rigorous fairness guarantees. The use of the SV for pricing personal data can be traced to~\cite{10.5555/1028128.1028156}, which studied the use of SV in the context of marketing survey, collaborative filtering, and recommendation systems. Recently, Jia et al.~\cite{DBLP:journals/pvldb/JiaDWHGLZSS19} formally studies the notion of fairness based on the SV and designed an algorithm that can calculate the SV more efficient based on Nearest Neighbor Algorithms (KNN). 
Although the interaction between data analytics and economics has been extensively studied, data security is often neglected, especially in the data evaluation process, which is fundamental in the data marketplace. 
Xu et al.~\cite{DBLP:journals/corr/abs-2012-06430} developed three methods for data appraisal: Norm of Parameter Gradients, Model Fine-tuning and Influence Functions, and proposed to perform data appraisal using MPC. However, their method cannot be applied to calculate SV. Azcoitia et al.~\cite{DBLP:journals/corr/abs-2012-08874} proposed a algorithm namely $\textit{Try Before You Buy}$ (TBYB) to evaluate data value before paying for them. However, it requires a sandbox to train the target model multiple times using different datasets. 



\section{Conclusion and future work}
\label{sec:conclusion}
In response to the increasing demand for protecting security during data transactions in the data marketplace, we propose a fair payment scheme suitable for the data marketplace while satisfying the demand for data valuation. We develop algorithms that are useful for data valuation while friendly be implemented via MPC circuit. To guarantee that the data to be paid is exactly the data evaluated before, we innovatively incorporate fair payment into the MPC circuit instead of using ZKCP. 
Since it is general, efficient, and also conceptually easy to understand, we believe it to have the potential to serve as the de facto method for real-world data marketplace platforms.
In the future, it would be promising to see further studies on reducing the number of pre-shared data with the buyer and improving the efficiency of the framework.


\bibliographystyle{IEEEtran}
\bibliography{references}

\end{document}